\documentclass[aps,prl,twocolumn,reprint,superscriptaddress,amsmath,amssymb]{revtex4-1}

\usepackage{graphics,color}
\usepackage{color,soul}
\usepackage{graphicx}
\usepackage{natbib}
\usepackage{amssymb,latexsym,amsmath}
\usepackage{verbatim}
\usepackage{rotating}
\usepackage{hyperref}

\newcommand{\miniscule}{\@setfontsize\miniscule{4}{5}}

\def\saychris#1{\emph{\bf \color{red} (CHRIS: #1)}}

\begin{document}
    
    \title{Eigenvector Centrality Distribution for Characterization of Protein Allosteric Pathways}
    
    \author{Christian F. A. Negre} 
    \thanks{Equally contributing authors} 
    \email{cnegre@lanl.gov}
    \affiliation{Theoretical Division, Los Alamos National Laboratory, Los Alamos, NM 87545}

    \author{Uriel N. Morzan} 
    \thanks{Equally contributing authors} 
    \email{uriel.morzan@yale.edu}
    \affiliation{Department of Chemistry, Yale University, P.O. Box 208107, New Haven, CT 06520-8107, and Energy Sciences Institute, Yale University, P.O. Box 27394, West Haven, CT 06516-7394}

    \author{Heidi P. Hendrickson} 
    \affiliation{Department of Chemistry, Yale University, P.O. Box 208107, New Haven, CT 06520-8107, and Energy Sciences Institute, Yale University, P.O. Box 27394, West Haven, CT 06516-7394}

    \author{Rhitankar Pal} 
    %
    \affiliation{Department of Chemistry, Yale University, P.O. Box 208107, New Haven, CT 06520-8107, and Energy Sciences Institute, Yale University, P.O. Box 27394, West Haven, CT 06516-7394} 
    
    \author{George P. Lisi} 
    %
    \affiliation{Department of Chemistry, Yale University, P.O. Box 208107, New Haven, CT 06520-8107, and Energy Sciences Institute, Yale University, P.O. Box 27394, West Haven, CT 06516-7394} 
    
    \author{J. Patrick Loria} 
    %
    \affiliation{Department of Chemistry, Yale University, P.O. Box 208107, New Haven, CT 06520-8107, and Energy Sciences Institute, Yale University, P.O. Box 27394, West Haven, CT 06516-7394}
    
    \affiliation{Department of Molecular Biophysics and Biochemistry, Yale University, New Haven, CT, United States.}
    
    \author{Ivan Rivalta} 
    \email{ivan.rivalta@ens-lyon.fr}
    \affiliation{\'Ecole Normale Sup\'erieure de Lyon, CNRS, Universit\'e Lyon 1, Laboratoire de Chimie UMR 5182, 46, All\'ee d'Italie, 69364 Lyon Cedex 07, France.} 
    
    \author{Junming Ho} 
    \affiliation{Agency for Science, Technology and Research, Institute of High Performance Computing, 1 Fusionopolis Way \#16-16 Connexis North, Singapore 138632.} 
    
    \author{Victor S. Batista} 
    \email{victor.batista@yale.edu}
    \affiliation{Department of Chemistry, Yale University, P.O. Box 208107, New Haven, CT 06520-8107, and Energy Sciences Institute, Yale University, P.O. Box 27394, West Haven, CT 06516-7394} 
    
    \begin{abstract}
        Determining the principal energy pathways for allosteric communication in biomolecules, that occur as a result of thermal motion, remains challenging due to the intrinsic complexity of the systems involved. Graph theory provides an approach for making sense of such complexity, where allosteric proteins can be represented as networks of amino acids.  In this work, we establish the eigenvector centrality metric in terms of the mutual information, as a mean of elucidating the allosteric mechanism that regulates the enzymatic activity of proteins. Moreover, we propose a strategy to characterize the range of the physical interactions that underlie the allosteric process. In particular, the well known enzyme, imidazol glycerol phosphate synthase (IGPS), is utilized to test the proposed methodology. The eigenvector centrality measurement successfully describes the allosteric pathways of IGPS, and allows to pinpoint key amino acids in terms of their relevance in the momentum transfer process. The resulting insight can be utilized for refining the control of IGPS activity, widening the scope for its engineering. Furthermore, we propose a new centrality metric quantifying the relevance of the surroundings of each residue. In addition, the proposed technique is validated against experimental solution NMR measurements yielding fully consistent results. Overall, the methodologies proposed in the present work constitute a powerful and cost effective strategy to gain insight on the allosteric mechanism of proteins.   
    \end{abstract}
    
    \maketitle
    
    Allostery is a ubiquitous process of physico-chemical regulation in biological macromolecules such as enzymes. 
    The fundamental step in the allosteric regulation is the binding of a ligand at a particular enzymatic site affecting the activity at a different and often very distant position of the protein. 
    While allosteric processes have long been of interest, especially due to their relevance in developing potent and selective therapeutics, the mechanism for energy transfer between allosteric sites remains poorly understood. 
    Thus, establishing a molecular level understanding of communication pathways between the physically distant enzymatic sites is crucial for the design of innovative drug therapies\cite{Csermely2013,Wagner2016} and protein engineering\cite{Goodey2008, Reetz2009, Ozbil2012}. 
    
    Recently, there have been significant efforts toward the development of computational tools to support, interpret and/or predict experimental evidences for elucidation of allosteric pathways in proteins \cite{Hawkins2004, Ming2005, Palumbo2006, Rivalta2012, VanWart2012,Wagner2016,Ribeiro2016,Plos-Blacklock-2014}. Network analysis has been extensively used in this context, by incorporating concepts and methodologies from graph theory into the realm of molecular dynamics simulations \cite{JPCB-Yaoquan-2014,JPCB-Yuzhen-2016,Biochem-sanjib-2016,JPCB-Lishi-2015,JCTC-VanWart-2012,JACS-palermo-2017,Chemrev-Guo-2016}
    For instance, community network analysis (CNA) has emerged as a powerful and increasingly popular approach to analyze the dynamics of enzymes and protein/DNA (and/or RNA) complexes and to detect possible allosteric pathways \cite{PONE-Li-2014,PNAS-Sethi-2009,SCIREP-Ricci-2016,Papaleo2012,David-Eden2008,Jiang2010,Szilagyi2013}. 
    
    In these network theory-based approaches, a protein is represented as a network consisting of a set of nodes, $n$ connected by edges, $m$. Usually, each amino acid is associated to a node  (typically positioned on the alpha carbon or the center of mass of the residue side chain). Depending on the physical property of interest, there are multiple quantities that can characterize the edges (i.e. the connections between nodes), such as the magnitude of the dynamical correlations \cite{Rivalta2012,Lange2006,Lange-2008}, the energetic coupling \cite{Savoie-2014}, or the spatial distance between residues \cite{Doshi-2016}. Given a network of N nodes,  its graph can be represented with an N $\times$ N adjacency matrix $\mathbf{A}$ whose elements $\mathbf{A}_{ij}$ are related with the strength of the physical interaction under consideration.
    
    One of the corner stones of network analysis is the concept of centrality, i.e.,  the relative importance of a node or clusters of nodes. Measures of centrality are crucial to identify the most influential nodes in a network. The importance is usually quantified by a real-valued function, related to a type of flow or transfer across the network (e.g., the amount of momenta transported by a given atom in a protein). There are many  measures of centrality characterizing slightly different aspects of the network. Probably the simplest of all is the degree, $k_i$, of each node, $i$, which is defined by the number and strength of the connections attached to it,

    \begin{equation}
        k_{i} = \sum_{j=1}^{n}{\mathbf{A}_{ij}} .
    \end{equation}

    The degree centrality (DC), provides a measure of the relative connectivity of each node within a network. A node that is well connected is expected to have a large ``influence'' on the graph. While the DC can provide useful information, it is not a true ``node-centrality'' as defined by Ruhnau,\cite{Ruhnau2000} and thus does not give a measure of centrality based on a fixed scale that allows comparisons between different graphs.
    
    An alternative definition is the betweenness centrality (BC), $b_i$, which provides a measure of how information can flow between nodes (or edges) in a network. The BC can be quantified as the number of times a node acts as a bridge along the geodesic (shortest) path between two other nodes,   

    \begin{equation}
        b_{i} = \sum_{st}{\frac{n_{st}^{i}}{g_{st}}} ,
    \end{equation}

    where $n_{st}^{i}$ is number of shortest paths between nodes $s$ and $t$ that pass though node $i$, and $g_{st}$ is the total number of shortest paths between nodes $s$ and $t$. The nodes with high BC have a large influence on the overall information passing, and hence, the removal of such nodes may disrupt the communication in the network. However, communication do not always take the shortest path, and hence, the BC may provide a misleading interpretation of the real relevance of each amino acid in the functional dynamics of a protein.    
    
    Somehow in between these two definitions of centrality (i.e. degree and betweenness centralities), the eigenvector centrality (EC) emerges as an alternative that takes into account both the number of connections of a given node and its relevance in terms of information flow.  The EC of a node, $c_i$, is defined as the sum of the centralities of all nodes that are connected to it by an edge, $\mathbf{A}_{ij}$, 
    \begin{equation}
        c_{i} = \epsilon^{-1} \sum_{j=1}^{n}{\mathbf{A}_{ij}c_{j}} ,
        \label{ec}
    \end{equation}
    therefore, $\mathbf{c}$ is the eigenvector associated to the eigenvalue $\epsilon$ of $\mathbf{A}$.  The EC is, hence, a measure of how well connected a node is to other well connected nodes in the network. Noteworthy, the EC serves as a measure of the connectivity against a fixed scale when normalized, and so it can be used to reliably compare different networks.\cite{Ruhnau2000} For example, the normalization becomes essential when analyzing differences between graphs, e.g., to study the pattern of centrality variation between the {\it apo} and {\it holo} states of a protein. 
    
\begin{figure}[h]
        \centering
        \includegraphics[width=8cm]{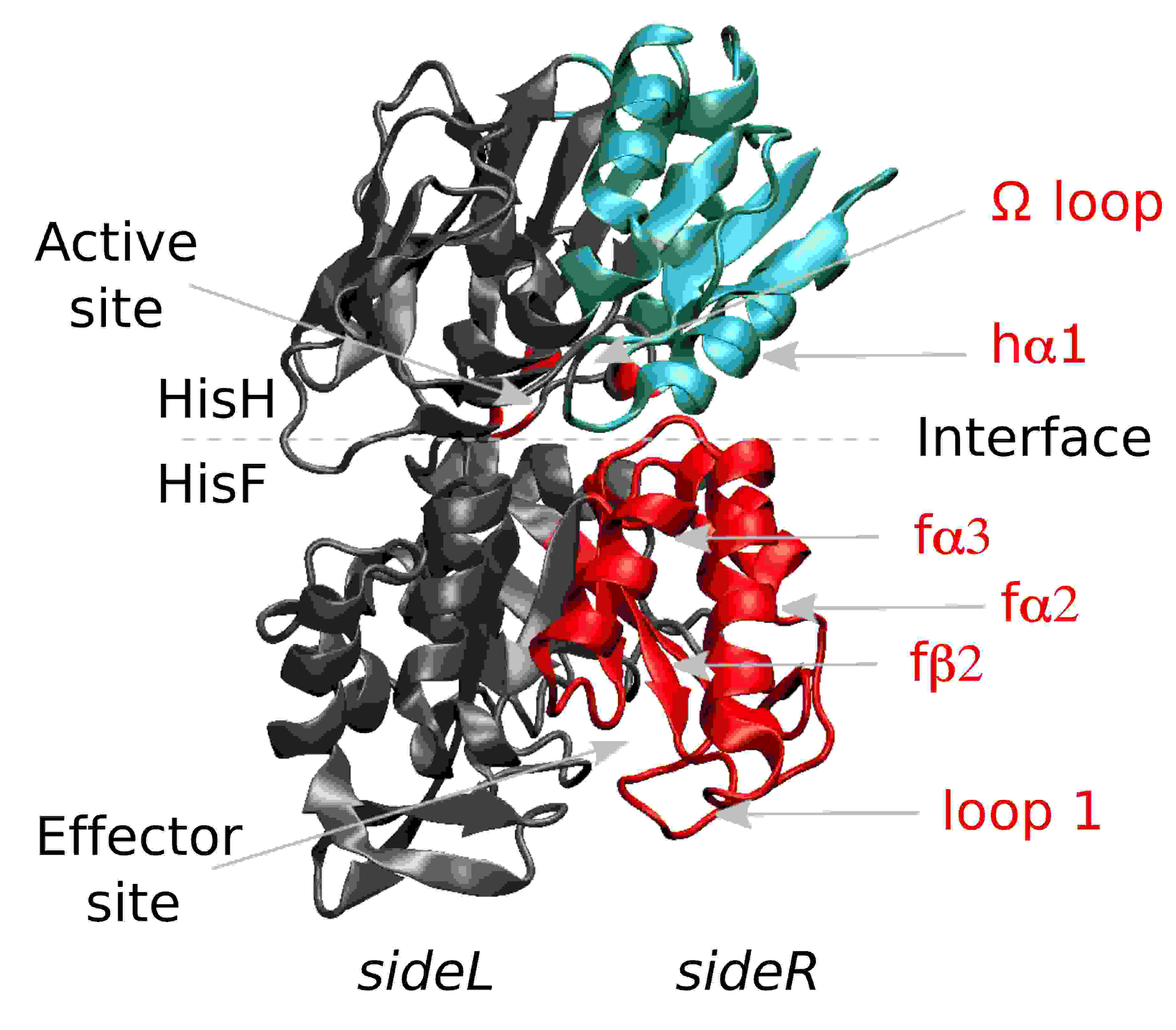}
        \caption{Molecular representation of IGPS. We have added labels for some key molecular features that are directly involved in the allosteric regulation. Communities \textbf{h2} (cyan) and \textbf{f3} (red) are also depicted.  
        }
        \label{igps}
    \end{figure}

    In the present work, we illustrate the potential of the EC measure to provide a molecular level characterization of the allosteric mechanism of enzymes. In particular, we focus on the prototypical case of the Imidazole Glycerol Phosphate Synthase (IGPS), a bacterial enzyme present in the amino acid and purine biosynthetic pathways of most microorganisms, making it an attractive target for antibiotic, pesticide, and herbicide development.\cite{Chaudhuri-2016} Structurally, IGPS is a tightly associated heterodimer (see Fig. \ref{igps}) in which each monomer catalyzes a different reaction: The {\it HisH} enzyme promotes the hydrolysis of glutamine (Gln) to produce ammonia, which diffuses to the {\it HisF} unit and reacts with the effector PRFAR to form imidazole glycercol phosphate (IGP). While Gln binding is unaffected by the presence of PRFAR, the hydrolysis of Gln is accelerated 5000-fold upon PRFAR binding through a mechanism that, for many years, has remained elusive \cite{Myers2003}. IGPS is thus a V-type enzyme and a model system to study non-cooperative allostery involving conformational changes.
    
    In a recent study \cite{Rivalta2012}, we have carried out a BC-based  community network analysis by optimizing the modularity function, to explore the underlying allosteric mechanism of this enzyme. We now present an alternative strategy, exploring the description of allostery provided by the EC as compared to the CNA based on optimal modularity (the connection between CNA and the EC is analyzed in detail in the SI). The results presented here are both complementary and fully consistent with our previous findings. Additionally, at variance with our previous CNA approach, the strategy proposed in this work allows to capture the long range contribution to the correlation pattern evidencing fundamental aspects of the allosteric behavior of IGPS. Therefore, the methodology presented here represents an ideal technique for the identification of mutation targets to inhibit or enhance the IGPS catalytic activity, opening the doors to a plethora of combined theoretical-experimental studies oriented to increase the control of its function and develop new alternatives for drug discovery.
    
    The present paper is organized as follows: We first summarize the method of CNA and results for reference \cite{Rivalta2012}. Next the method of EC is introduced and applied to the IGPS systems. Results are discussed and compared with CNA. Correlation matrices are obtained from the same trajectories and following the same protocol as in reference \cite{Rivalta2012} and \cite{Rivalta2012SI}.
    
    \section{Community Network Analysis}
    
    Consider a protein residue network where each node represents the $\alpha$-carbon of an amino acid in the protein, and each edge represents the dynamical correlation between the two residues (nodes) it connects. The latter can be quantified using the generalized correlation coefficients, based on the mutual information (MI) between two residues $\textbf{r}_{MI}[\textbf{x}_i, \textbf{x}_j]$ \cite{Lange2006}:
    \begin{equation}
        \textbf{r}_{MI}[\textbf{x}_i, \textbf{x}_j]= \left(1-\exp{\left(-\frac{2}{3}\textbf{I}[\textbf{x}_i, \textbf{x}_j]\right)}\right)^{1/2}
    \end{equation}
    where the fluctuation or atomic displacements vectors $\textbf{x}_k$ are computed from a molecular dynamics (MD) simulations. For clarity, we have kept the original notation used in references \cite{Lange2006,Rivalta2012,Rivalta2012SI}, where a detailed explanation on the calculation of the generalized correlation coefficients can be found. 
    
    The mutual information between the two residues is computed as: 
    \begin{equation}
        \textbf{I}[\textbf{x}_i, \textbf{x}_j] = H[\textbf{x}_i] + H[\textbf{x}_j] - H[\textbf{x}_i, \textbf{x}_j] ,
    \end{equation}
    where 
    %
    \begin{equation}
        H[\textbf{x}_i] = -\int p[\textbf{x}_i] \ln(p[\textbf{x}_i])  d\textbf{x}_i , 
    \end{equation}
    \begin{equation}
        H[\textbf{x}_i, \textbf{x}_j] =-\int\int p([\textbf{x}_i, \textbf{x}_j]) \ln(p([\textbf{x}_i ,\textbf{x}_j]) d\textbf{x}_i d\textbf{x}_j ,
    \end{equation}
    %
    are the marginal and joint Shannon entropies respectively, obtained as ensemble averages over the atomic displacements $(\textbf{x}_i, \textbf{x}_j)$, with marginal and joint probability distributions $p[\textbf{x}_i]$ and $p[\textbf{x}_i,\textbf{x}_j]$ computed over thermal fluctuations sampled by molecular dynamics simulations of the system at equilibrium. The coefficient $\textbf{r}_{MI}$ ranges from zero for  uncorrelated variables, to 1 for fully correlated variables.
    
    The protein graph connectivity is then built excluding direct connections of first nieghbors (in amino acid sequence) and according to two cutoffs: two nodes are considered connected if the distance between their  $\alpha$-carbons is within a distance cutoff (generally 4-6 \AA) for a certain percentage of the MD trajectories (percentage cutoff, usually 65-85 \%). 
    The distances between all the connected nodes ($i,j$) in the graph topology define a matrix of elements $\textbf{w}_{ij}^{(0)}$  obtained from $\textbf{r}_{MI}[\textbf{x}_i, \textbf{x}_j]$, according to: 
    \begin{equation}
        \textbf{w}_{ij}^{(0)}=-\log[\textbf{r}_{MI}[\textbf{x}_i, \textbf{x}_j]] ,
    \end{equation}
    setting the $\textbf{w}_{ij}$ distance to infinity (in practice to extremely large values) when two nodes are not connected, as defined by the connectivity rules. The Floyd-Warshall algorithm \cite{Floyd1962} is then used to determine the matrix of minumum distance (maximum correlation), $\textbf{w}_{ij}^{(\text{M})}$ considering direct distances as well as up to N possible intermediate residues mediating indirect communication pathways (where N is the total number of residues in the system). The total number of residues for the IGPS case is N$=454$.
    
    The edge-betweenness matrix with elements $\textbf{b}_{ij}$ is defined as the number of shortest paths that include edge ($m_{ij}$) as one of its communication segments. In other words, the edge-betweenness matrix is an estimation of the information ``traffic'' passing through the edge connecting residues $i$ and $j$ in the network. The edge-betweenness matrix is then used for partitioning the network into communities according to the Girvan-Newman algorithm which is based on maximizing the modularity $Q$ measure \cite{Girvan2002,Newman2006}. Details of the computation of the communities structure based in the maximum modularity from the generalized correlation matrix can be found in references \cite{Rivalta2012,Rivalta2012SI}.
    
    Figure \ref{igps} shows the two most important communities \textbf{h2} (cyan) and \textbf{f3} (red) projected into the residue space of the IGPS in the {\it apo} state as determined in \cite{Rivalta2012}. Secondary structural element of \textbf{h2} involves h$\beta1$, h$\beta2$, h$\beta3$, h$\beta4$, h$\beta11$, h$\alpha1$, h$\alpha2'$ and $\Omega$-loop. Secondary structural element of \textbf{f3} instead involves f$\beta1$, f$\beta2$, f$\beta3$, h$\beta7$, h$\beta8$, f$\alpha1$, f$\alpha2$, f$\alpha3$, h$\alpha4$ and Loop1.  
    
    We have previously showed that the correlation between communities \textbf{h2} and \textbf{f3} is enhanced (with larger inter-betweenness) after PRFAR binding. Furthermore, it was shown that the explanation for this enhancement relies on the increase in the frequency of an interdomain motion at the dimeric interface ({\it HisH}-{\it HisF}) upon the binding of PRFAR. This was described as a low-frequency inter-domain breathing motion that allows for fluctuations between two states (open and closed IGPS heterodimer) that are accessible at thermal equilibrium in both the {\it apo} and PRFAR complexes. Disruption of this breathing mode with drug-like compounds was recently suggested as a method for inhibiting the allosteric mechanism \cite{Rivalta2016}.
    
    The recognition of the local interactions that determine variations in the breathing motion (and, thus, in the \textbf{h2}-\textbf{f3} inter-communities correlations) has been performed by detailed comparative analysis of chemical interactions along the MD trajectories of {\it apo} and PRFAR-bound IGPS complexes \cite{Rivalta2012}. In particular, it was observed  that PRFAR binding affects specific hydrofobic interactions in Loop1 and f$\beta$2 (in {\it HisF}), altering salt-bridge formations at the surface exposed f$\alpha$2, f$\alpha3$ and h$\alpha1$ helices (at the {\it HisF}/{\it HisH} interface) that, in turn, determine modification of the breathing motion and of the hydrogen bonding network between the Omega-loop and the oxyanion strand nearby the {\it HisH} active site. Thus, among the secondary structure elements of communities \textbf{h2} and \textbf{f3}, the following elements have been retained as allosteric pathways: Loop1, f$\beta2$, f$\alpha2$, f$\alpha3$, h$\alpha1$ and $\Omega$-loop (indicated with red labels in Figure \ref{igps}). The active allosteric role of some of these residues has been recently proved by single-site mutation experiments \cite{PNAS-Lisi-2017}.
    
    The CNA provides an introspection tool for visualizing the most important transformation induced by the allosteric effector in a coarse-grained fashion, allowing easy detection of effector-driven changes in the overall inter-communities information flows. However, we have showed that to recover direct information on allosteric pathways, a detailed analysis of the MD trajectory is still necessary \cite{Rivalta2012}. Therefore, CNA can successfully assist the tedious allosteric pathways detection by indicating major network changes due to the effector binding but it cannot provide an easy detection and immediate visualization of the sequence of amino acids involved in the allosteric-to-active site signal propagation.  Here we show that a comparative EC approach on the other hand, can provide fast detection of allosteric nodes and easy interpretation of the signal pathways ``activated'' by the effector binding.
    
    \section{Eigenvector Centrality Analysis}
    
    Lets define the adjacency matrix as follows:
    
    \begin{equation}
        \mathbf{A}_{ij} = \left\{\begin{array}{clrr}
            0, & \mathrm{if} \quad i=j  \\
            {\bf r}_{MI}[\textbf{x}_i, \textbf{x}_j] \exp(-\frac{d_{ij}}{\lambda}) &  \mathrm{if}  \quad  i\neq j .
        \end{array}\right.
        \label{adj}
    \end{equation}
    
    Just as in the CNA approach, here each node of the graph corresponds to the $\alpha$-carbon of an amino acid residue and the off-diagonal elements of $\mathbf{A}$ are the weights associated to every edge. Additionally, an exponential damping factor with a length parameter $\lambda$ has been introduced to expression \ref{adj}. This parameter can be adjusted to control the locality of the correlations under consideration based on the average distance between residues ($d_{ij}$). This means that if $\lambda$ is short enough, the correlation between residues that are far away from one another will be disregarded and the effect of the locality in the allosteric pathway will be revealed. On the other hand, if $\lambda$ is set to a very large value, all correlations, including those between residues separated by long distances, will be accounted for (i.e. $\lambda \rightarrow \infty$, $\mathbf{A}_{ij} = \textbf{r}_{MI}[\textbf{x}_i, \textbf{x}_j]$ $\forall \; i\neq j$). By adopting such damping factor, we obtain a two-fold benefit for the EC analysis: i) by setting reasonably small damping values we could mimic the distance cutoff employed in the CNA and we can then fairly compare EC and CNA results; ii) comparison of EC values at various damping distances provides direct information on the role of long-range correlations in allosteric pathways. This will be discussed in further detail in the last section. 
    
\begin{figure}[h]
        \centering
        \includegraphics[width=9cm]{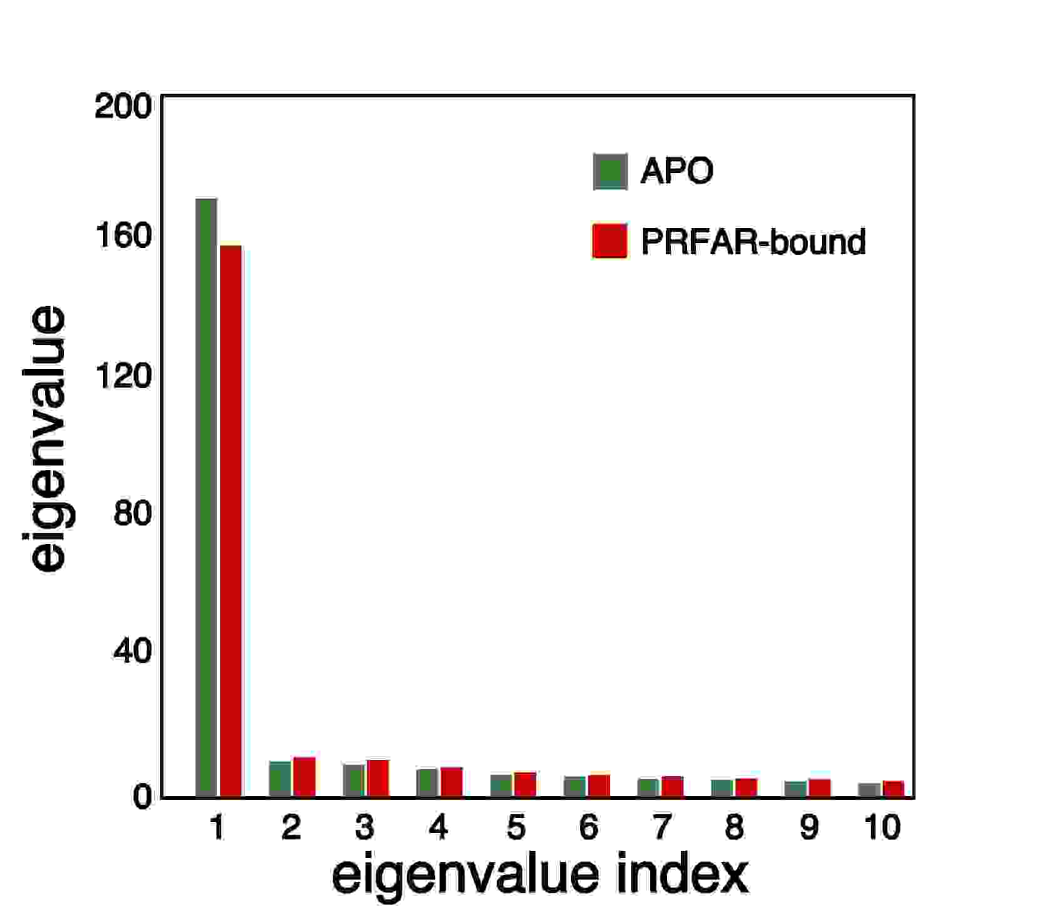}
        \caption{Largest 10 eigenvalues obtained form the adjacency matrix defined by \ref{adj} (in the limit of $\lambda \to \infty$) for the {\it apo} (green) and PRFAR bound (red) IGPS. \saychris{isn't this part of the SI?}}
        \label{EIGENVAL}
    \end{figure}

    As mentioned in the introduction, the eigenvector centrality (EC) measurement arises from an eigendecomposition of the adjacency matrix, $\mathbf{A} \bf{c}=\epsilon \bf{c}$, where $\textbf{c}$ is the vector containing the centralities $c_i$ for each node $i$ and $\epsilon$ is the associated eigenvalue.  Therefore, there is a set of $N$ solutions to this eigenvalue problem, being $N$ the number of $\alpha$-carbon atoms in the protein. However, we will rely here on the assumption that the functional dynamics of the protein can be assigned to the major collective mode of correlation. Consequently, the eigenvectors associated to the remaining eigenvalues will be neglected. The election of this leading eigenvector as the principal component of the correlation pattern can be formally justified considering that the  adjacency matrix $\mathbf{A}$ defined by equation \ref{adj}, has the following mathematical properties: (\textbf{i}) $\mathbf{A}_{ij}=\mathbf{A}_{ji}$ $\forall$ $i,j$; and (\textbf{ii}) $0 \leq \mathbf{A}_{ij} \leq 1$ $\forall$ $i,j$ . Hence, uniqueness of the definition of the eigenvector centrality is ensured by the Perron-Frobenius theorem which states that any symmetric matrix (property \textbf{i}) with non-negative entries (property \textbf{ii}) has a unique largest real eigenvalue (see SI). To illustrate the practical consequence of this theorem in the case of $\it apo$ and PRFAR bound IGPS, Figure \ref{EIGENVAL} shows that there is almost two orders of magnitude separating the highest eigenvalues from the remaining ones.  
    
    Based on this definition, the EC values $c_i$ can be computed by diagonalizing matrix $\mathbf{A}$ and keeping the eigenvector $\textbf{c}$ that corresponds to the maximum eigenvalue $\epsilon$. The power method \cite{watkins2010} is an alternative to matrix diagonalization which is computationally more efficient and would be more appropriated for large systems. The information encoded on the resulting eigenvector $\textbf{c}$ reveals the importance of the nodes for the whole connectivity of the network. The nodes with the highest centralities will act as the principal ``channels'' for momentum transmission across the protein. This strategy has been applied as a means of visualizing dynamical phenomena in other domains of science \cite{Jimenez2017}.
    
    As the set of eigenvectors of $\mathbf{A}$ is orthonormal, the sum of all the squared centralities is one  ($\sum_i c_i^2 = 1$). The latter plus the fact that the centralities are positive suggests that the squared centralities $c_i^2$ could be interpreted as the probability for a signal to pass through node $i$ \cite{Jimenez2017}. 
    The eigenvalue $\epsilon$, in turn, gives a measure of the network degree of connectivity. At $\lambda \to \infty$ (no exponential damping), the values of $\epsilon$ are 166.8 and 154.0 for {\it apo} and PRFAR-bound respectively. This indicates that the system experiences an overall decrease of correlation as a consequence of PRFAR binding as previously suggested by inspecting the correlation matrix \cite{Rivalta2012}. Moreover, our solution NMR spectroscopic measures characterizing the conformational exchange ($k_{ex}$) for numerous amino acids in {\it HisF} domain, indicate that nearly every residue increases its flexibility upon PRFAR binding \cite{Lisi-struct-2016}. This increase in flexibility is translated into an effective reduction of the intermolecular connectivities, and hence, results fully consistent with the predicted drop in the overall correlation.

\begin{figure}[h]
        \centering
        \includegraphics[width=8cm]{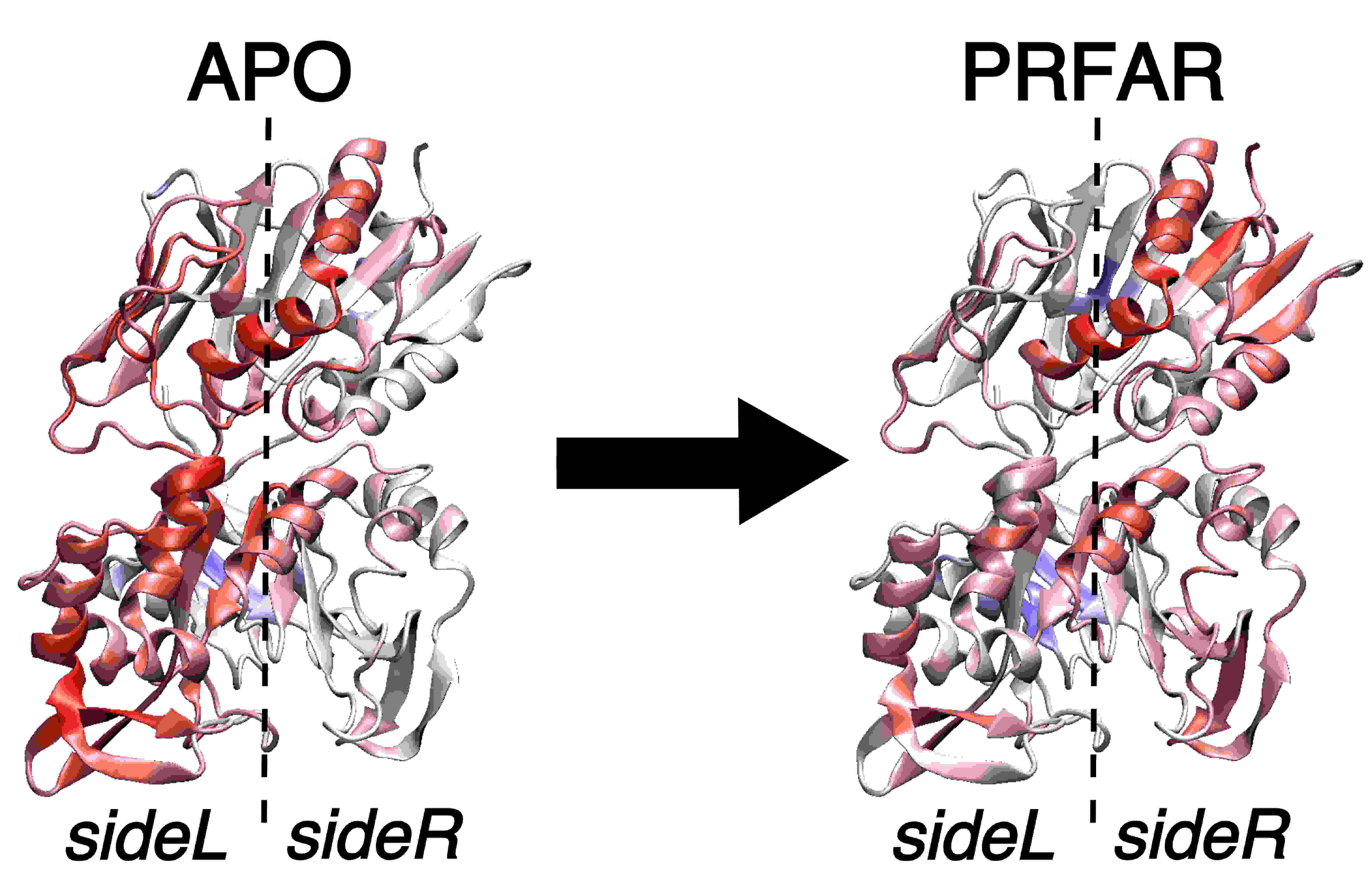}
        \caption{Computed centrality values for both APO and PRFAR-bound IGPS. The color scale goes from blue ($c = 0.0$) to red (max values of $c$).
        }
        \label{apo2prfar}
    \end{figure}

    The EC values for each node can be easily visualized in the protein structure (Figure \ref{apo2prfar}), displaying the $c_i$ coefficients for each amino acid with a color scale from white (zero centrality) to red (maximum centrality). In all the cases, a renormalization of the centrality values was applied for plotting purposes (See SI). Figure \ref{apo2prfar} shows the values of $\mathbf{c}$ for both {\it apo} and PRFAR-bound IGPS proteins, as computed by setting the damping distance to infinity. Noteworthy, the subgraph composed by the most important nodes in the network changes dramatically with the effector binding, highlighting the connection between the EC distribution and the momentum transport pathway.  As indicated in Figure \ref{apo2prfar}, the highest EC values shift collectively from {\it sideL} to {\it sideR} of the IGPS PRFAR binding.  This variation of the relative EC distribution evidences a change in the correlation pattern that is in agreement with our previous analysis and consistent with the enhancement in the betweenness of  \textbf{h2}-\textbf{f3} pair of communities  \cite{Rivalta2012}.
    
    The methodology introduced above somehow resembles the well known essential dynamics (ED) scheme in which the global trajectory of a system analyzed in terms of its major collective modes of fluctuation.\cite{ED-Amadei1993,ED-Hayward2008,ED-meyer2006,ED-prot2013} These modes -- usually called essential modes -- are obtained by diagonalizing the covariance matrix defined as  
    \begin{equation}
        \mathbf{C}_{ij}= \langle(\mathbf{x}_i(t)-\langle\mathbf{x}_i(t)\rangle) (\mathbf{x}_j(t)-\langle\mathbf{x}_j(t)\rangle)\rangle.
        \label{covariance}
    \end{equation}

\begin{figure}[h]
        \centering
        \includegraphics[width=8.5cm]{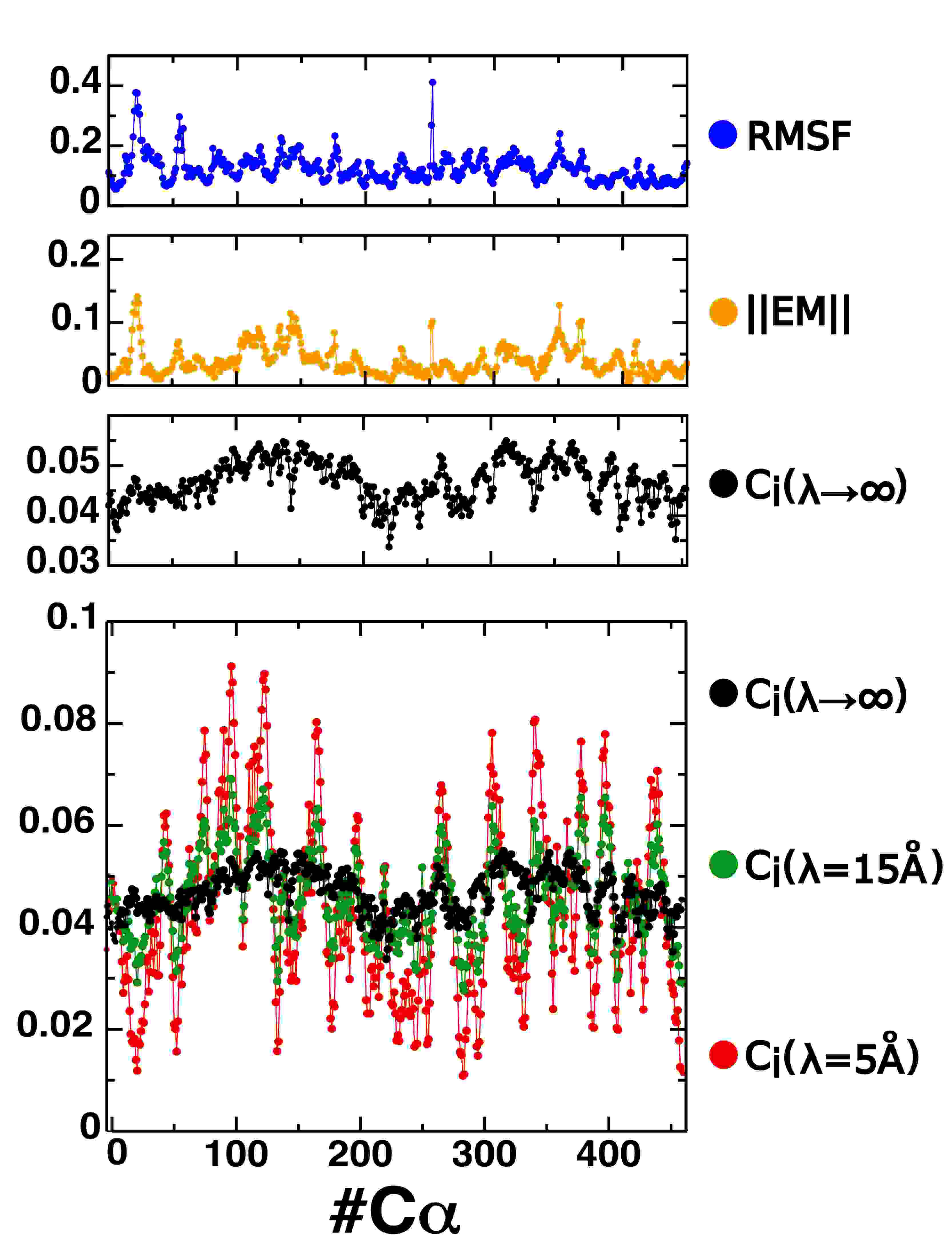}
        \caption{(Upper Panel) Comparison between the Euclidean norm of the elements of the first essential mode associated with each C$_\alpha$ (orange line), the centrality coefficients obtained from the first eigenvector of the adjacency matrix defined in equation \ref{adj} with $\lambda \to \infty$ (black line), and root mean square fluctuation per residue (RMSF) (blue line). (Lower Panel) Effect of the length parameter in the exponential damping factor of the adjacency matrix defined in equation \ref{adj}. Values of $\lambda = 5 $ \AA$, 15 $ \AA$ $ and $ \lambda \to \infty$ are depicted in red, green and black respectively.}
        \label{EC-ED}
    \end{figure}

    Normally, despite not being formally guaranteed, it is observed that the protein dynamics is dominated by a few essential modes. Therefore, this scheme also provides a way to obtain eigenvector coefficients that reveal the relevance of each node in the overall behavior of the network. Nevertheless, the measure of relevance can have several meanings, in particular the upper panel of Figure \ref{EC-ED} $ $ shows that the nature of the eigenvector coefficients obtained from the first essential mode (the one associated to the highest eigenvalue) is qualitatively different from that of the EC coefficients. There are two main reasons that justify this difference: (i)  while in the latter case the generalized mutual information matrix is only a measure of the dynamical correlation between pairs of nodes, in the former case the covariance matrix is both a measure of correlation and the amount of fluctuation. (ii) On the other hand, the covariance measure fails to account for non-colinear correlations. The first observation is consistent with the fact that the behavior of the essential mode coefficients (orange line, upper panel) is quite similar to the root mean square fluctuation per residue (blue curve, upper panel). Therefore, this analysis illustrates that the ED and the EC extracted from the mutual information are two complementary methodologies that provide a different insight on the systems dynamics. In particular the technique presented in this work constitutes a powerful alternative to analyze allosterism because it isolates the principal component in terms of the correlation and not in terms of flexibility as in the case of essential dynamics.
    
    The lower panel of Figure \ref{EC-ED} shows the effect of the length parameter $\lambda$ defined in expression \ref{adj}. In the limit of $\lambda \to \infty$ the off-diagonal elements of the adjacency matrix become equivalent to the generalized correlation function for each pair of nodes. The centrality coefficients obtained in this way exhibit a smooth variation. In contrast, when $\lambda$ is short enough, only the local components of the correlations survive and the centrality coefficients reveal the relevance of each residue in terms of its dynamical correlation with neighboring aminoacids. In this context, the exponential damping appears as a strategy to elucidate the \textit{correlation paths} triggered by short range molecular interactions, thus providing a physically relevant description of the momentum transfer within the protein residue network.
    
    \section{Centrality variation triggered by effector binding}
    
\begin{figure}[h]
        \centering
        \includegraphics[width=9cm]{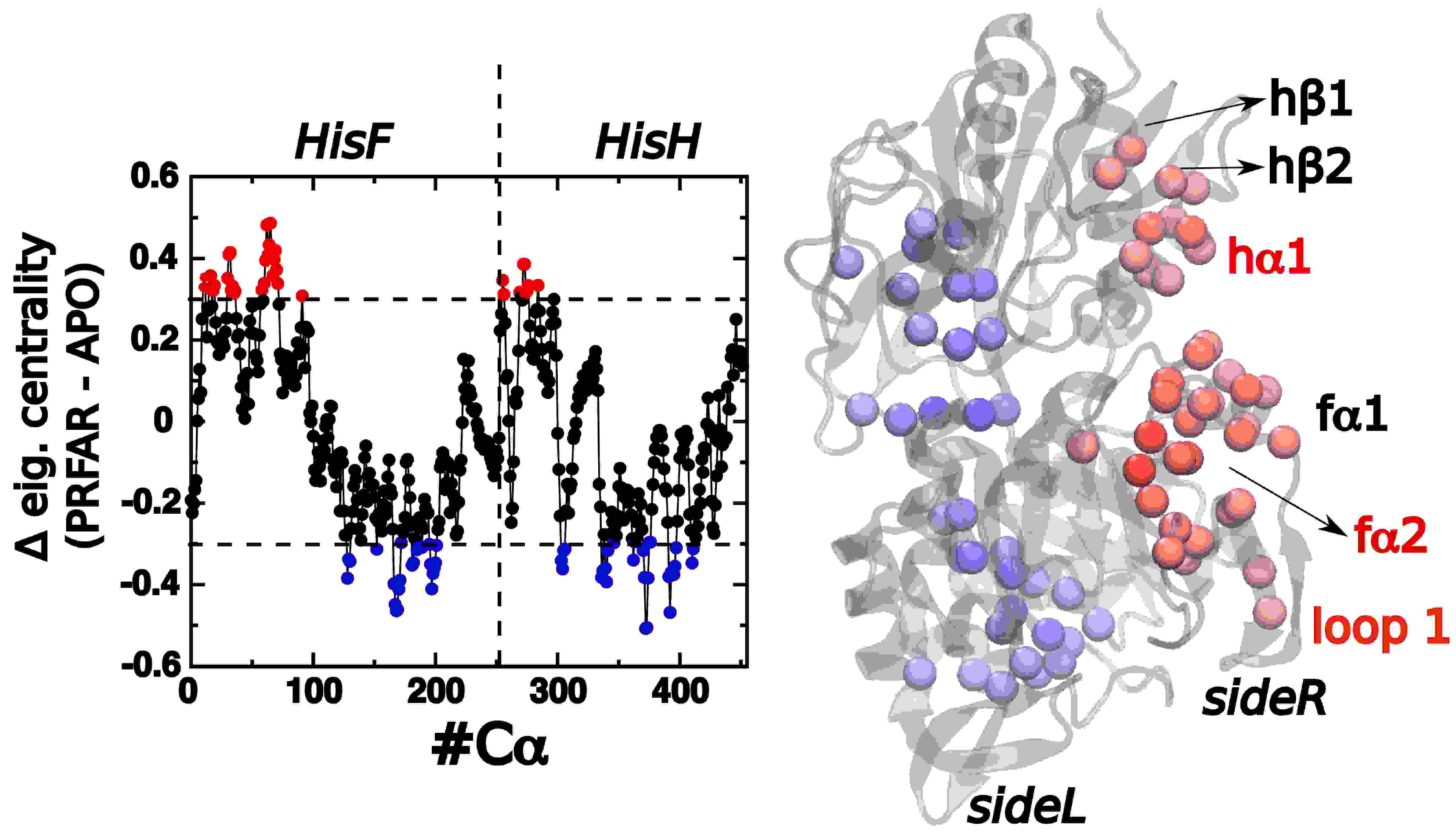}   
        \caption{Centrality differences (PRFAR-bound - APO) for an exponential damping $\lambda = 5$ \AA\, as a function of the residue index (a) and plotted on top of the protein representation (b). Red and blue values are regions that respectively gain and lose centrality upon PRFAR binding.}
        \label{comp}
    \end{figure}

    In order to highlight the changes in the EC distribution caused by the binding of the effector PRFAR (see Figure \ref{apo2prfar}), we have examined the EC differences associated to PRFAR binding ($c_i^{PRFAR} - c_i^{APO}$) for each residue $i$. Figure \ref{comp} shows that there is significant redistribution of the EC values upon PRFAR binding. 
    Two protein regions feature increased centralities, namely residues around 5-100 (in HisF) and around 254-330 (1-46 in HisH), involving the f$\alpha1$, f$\alpha2$, loop1  and h$\alpha1$ fragments. Connections between the loop1 and $\Omega$-loop are hence established after PRFAR is bound to IGPS as suggested in Ref. \cite{Rivalta2012} and as clearly depicted in the centrality differences analysis presented in Figure \ref{comp}.

    Previous studies have suggested the existence of two dynamically differentiated sides in IGPS, i.e. left and right or {\it sideL} and {\it sideR} respectively \cite{Rivalta2012,Rivalta2016} (Figure \ref{comp}).  Detailed inspection of MD trajectories have suggested that the allosteric signal propagates through {\it sideR}. Noteworthy, in agreement with that observation, Figure \ref{comp} shows that the binding of the effector PRFAR causes an increase in the centrality values of {\it sideR} amino acids. Moreover, the pattern shown by the centrality distribution allows to clearly identify the two sides of IGPS, confirming our previous hypothesis.     
    
    Importantly, the residues identified by this analysis are perfect targets for mutations that may have a major impact on IGPS catalytic behavior. In particular, helix h$\alpha1$ appears as a specially promising and unexplored fragment for site directed mutagenesis experiments oriented to refine the control of IGPS activity. 
    
    In addition, instead of focusing on the nodes that are important {\it per se}, another criteria that can be relevant to guide mutagenesis efforts is to focus on the ``neighborhood'' of those nodes. This sort of modification may play a more subtle role on altering the proteins activity, which can be potentially relevant for applications like drug discovery in which the desired effect comes from disrupting the environment of key residues in the protein. A strategy to obtain this {\it neighborhood centrality} measure is to subtract the degree centrality (DC) coefficients from the original EC values:
    
    \begin{equation}
        c'_{i} = \epsilon^{-1} \sum_{j=1}^{n}{A_{ij}c_{j}} - \sum_{k=1}^{n}{A_{ik}} .
        \label{neighbor-cent}
    \end{equation}
    
\begin{figure}[h]
        \centering
        \includegraphics[width=9cm]{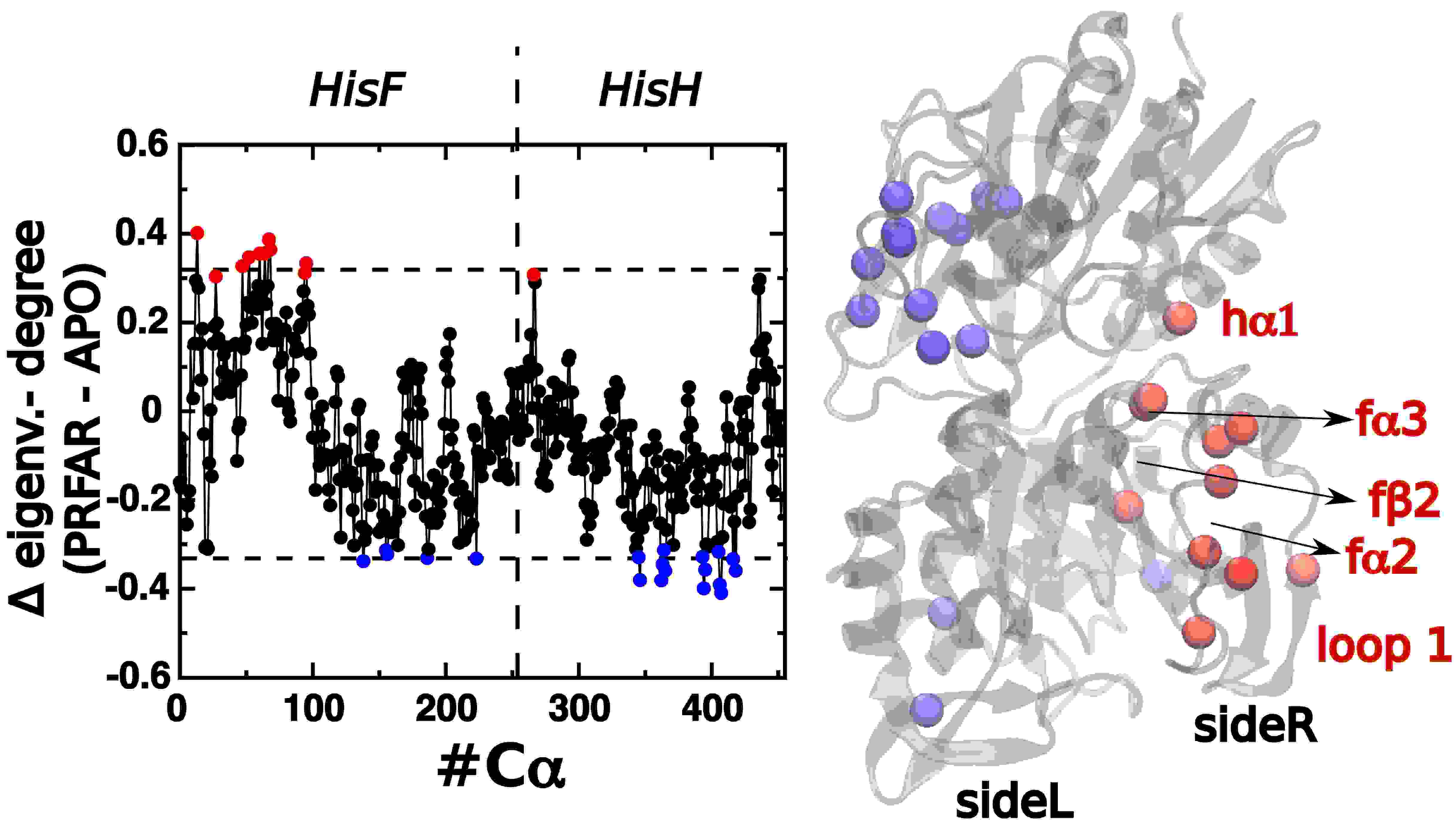}   
        \caption{Difference between EC and DC, $c'_{i}$, for the PRFAR binding process (PRFAR-bound - APO) for an exponential damping of $\lambda = 5$ \AA\, as a function or the residue index (a) and plotted on top of the protein representation (b). Red and blue values are regions that respectively gain and lose of centrality with central aminoacids upon PRFAR binding.}
        \label{comp-neighborhood}
    \end{figure}

    Figure \ref{comp-neighborhood}  illustrates the measurement of the $c'_{i}$ coefficients associated to the transition between the APO and PRFAR bound states (i.e. $c'_{i}=c'_{i}(PRFAR) - c'_i(APO)$). This analysis highlights residues fN14, fV48, fR59, fT61, fL65, fQ67, fV69, fR95,  fG96 and hN14 as the ones neighboring the aminoacids with a large increase of centrality upon PRFAR binding. With the exception of residues fT61, fL65 and fV69, all the aminoacids pointed out by this measurement coincide with the ones that have the larger PRFAR induced EC variation.  Remarkably, single-point mutation on residue fV48 and fN98 (which is in the vicinity fG96) have shown to have a dramatic effect on the PRFAR-induced activation of IGPS catalytic activity  \cite{PNAS-Lisi-2017}. On the other hand, the relevance of fV48 as part of the hydrophobic cluster in f$\beta$2 and fE67 and fR95 as part of the surface salt-bridge network at f$\alpha$2/f$\alpha$3 has been indicated by detailed MD trajectories inspections while here it is rapidly detected by the comparative EC analysis.
    
    \section{The locality factor} 
    
\begin{figure}[h]
        \centering
        \includegraphics[width=8cm]{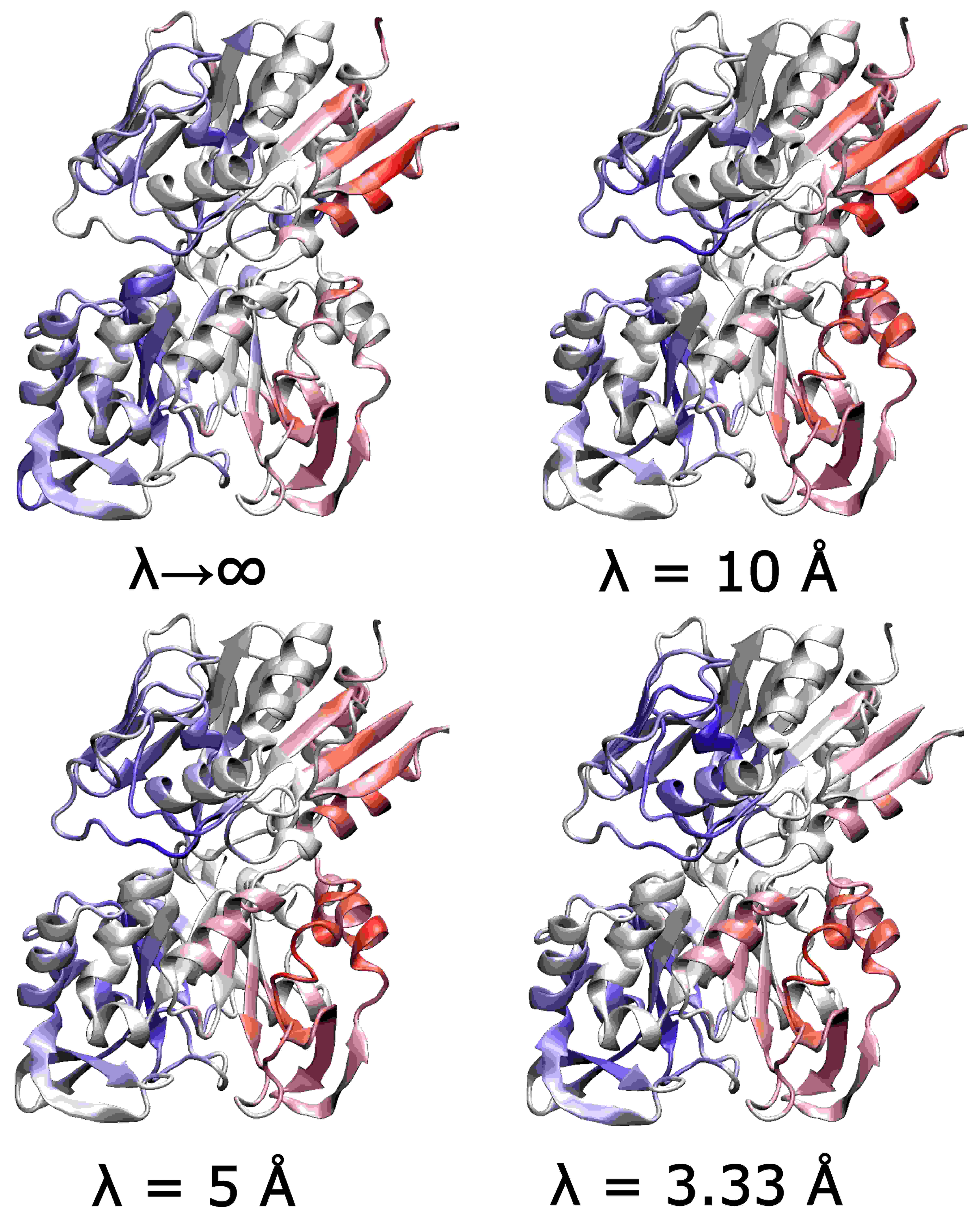}
        \caption{Centrality differences (PRFAR-bound - APO) for different values of         $\lambda$. Regions in red and blue correspond to gains and lose of     centrality respectively.}
        \label{dump}
    \end{figure}

    In order to further analyze the impact of the locality factor in the overall centrality distribution, Figure \ref{dump} shows the calculated EC coefficients at different values of $\lambda$. Importantly, adjusting the damping parameter down to $\lambda = 3.3$ \AA\, does not seem to have a significant effect on the overall trend of the EC differences between {\it apo} and PRFAR-bound IGPS. The same allosteric pathway for IGPS is revealed whether or not we include the correlations between residues separated by long distances. Moreover, the {\it sideL}/{\it sideR} structure is maintained at all $\lambda$'s. This implies that short range correlations dominate the protein dynamics, and hence the residue-to-residue effect is the main mechanism that underlies the momentum transmission in IGPS. Another important point to note is the fact that the disruptions of the centrality values disappear upon the application of the locality factor recovering the smoothness of a residue-to-residue short range transmitted signal.
    
    The average C$_\alpha$-C$_\alpha$ distance is around 3.8 \AA, therefore when the value of $\lambda < 4$ \AA, the correlation matrix becomes almost diagonal (see SI), and the key EC trend is most likely masked by numerical errors.
    
    \begin{figure}[h]
        \centering
        \includegraphics[width=5cm]{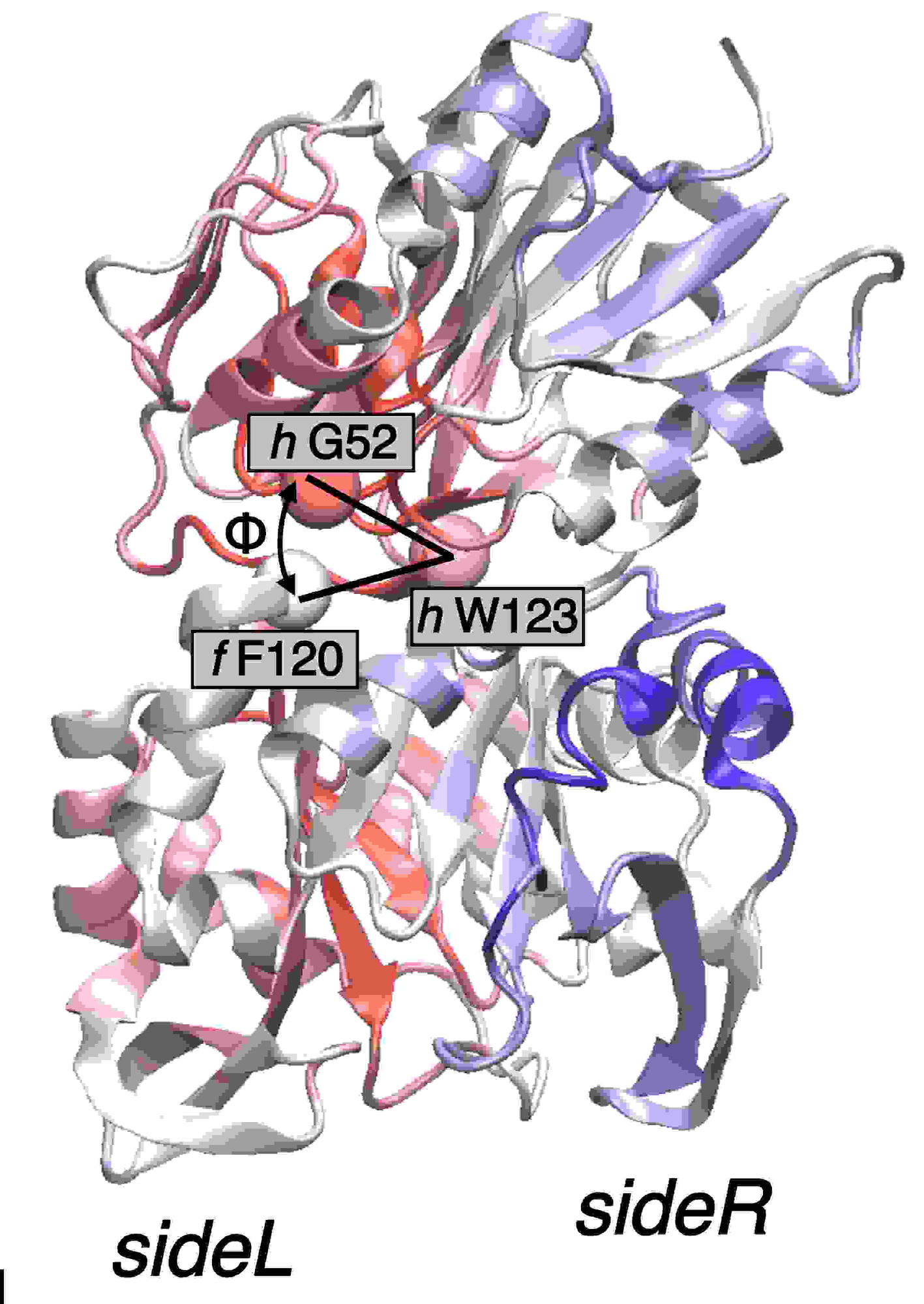}   
        \caption{Variation in the PRFAR-induced centrality coefficients caused by the application of the locality factor ($\lambda=5$ \AA). Red to blue scale characterizes a gain or loss of centrality respectively upon the application of the locality factor. }
        \label{comp-distance}
    \end{figure}
    
\begin{figure}[h]
        \centering
        \includegraphics[width=9cm]{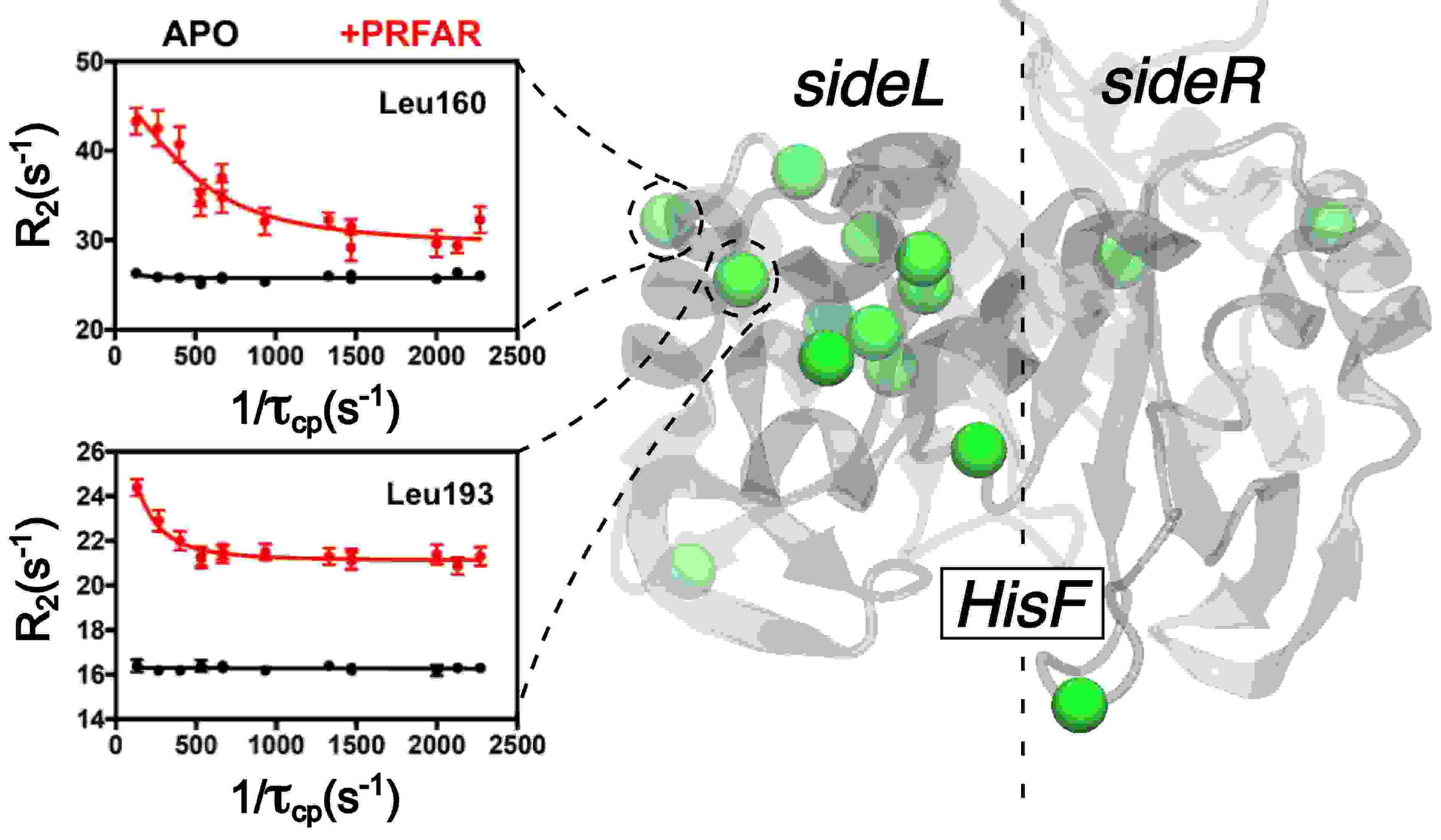}   
        \caption{NMR relaxation dispersion experiments characterizing the PRFAR-induced millisecond motions in {\it HisF} subunit of IGPS. The right panel highlights the residues that show the highest variation on their relaxation-dispersion profile upon IGPS binding. The left panel shows two representative relaxation dispersion curves for residues Leu160 (upper panel) and Leu193 (lower panel) in the APO and PRFAR bound states (black and red respectively).  }
        \label{NMR}
    \end{figure}

    As discussed above, by introducing the locality factor $\lambda$ it is possible to select from the overall motion of the system the correlations arising exclusively from physical interactions whose range are below certain distance threshold. On the other hand, despite  having shown that the resulting short range component is the one that dominates the overall correlation pattern, it is possible to analyze the nature of the long range contribution. Figure \ref{comp-distance} introduces a measurement of the long range component of the PRFAR induced EC coefficients computed as:
    
    \begin{equation}\label{xx}
        \begin{split}
            d_i^{\lambda_0}& =[c_{i}^\text{PRFAR} - c_i^\text{APO}]_{\lambda \to \infty} -[c_{i}^\text{PRFAR} - c_i^\text{APO}]_{\lambda = \lambda_0}\\
            & = [c_{i}^{\lambda \to \infty} - c_i^{\lambda=\lambda_0}]_{\text{PRFAR}} -[c_{i}^{\lambda \to \infty} - c_i^{\lambda=\lambda_0}]_{\text{APO}} ,\\
        \end{split}
    \end{equation}
    
    for $\lambda_0=5,10 \text{ and } 20$ \AA $ $ (panels A, B and C respectively) \saychris{Panels are missing?}.   Remarkably, the long range $d_i$ distribution also preserves qualitatively the {\it sideL}/{\it sideR} structure, but in this case the trends are inverted with respect to the short range picture, and the largest increase in the long range centrality coefficients upon PRFAR binding is mainly located on {\it sideL}. This is consistent with the presence of an interdomain ``breathing'' motion, as
    previously reported \cite{Rivalta2016,Rivalta2012} (Figure \ref{comp-distance}.A, dashed black lines forming an angle $\phi$). The large structural (long range) rearrangement associated to this motion increases its frequency upon PRFAR binding almost four times \cite{Rivalta2016}. Consequently, the highest gain of long range correlation that occurs mainly in {\it sideL} can be assigned to this low frequency motion. In agreement with this, our solution NMR relaxation dispersion experiments show that the PRFAR-induced millisecond motions are primarily located on {\it sideL} (Figure \ref{NMR}), which supports the existence of a large motion with maximum amplitude on {\it sideL} as determined by the long range centrality analysis. Furthermore, {\it sideL} of subunit {\it HisF} appears more static with weaker effectors than PRFAR \cite{Lisi-struct-2016}, suggesting that this breathing motion might be determining the allosteric activation of IGPS in some extent.  But more generally, the NMR study presented in Figure \ref{NMR} provides an experimental proof for the presence of the {\it sideL}/{\it sideR} structure predicted by the EC analysis, in which the two sides of IGPS display clear differences in terms of their dynamical features. 
    
    Interestingly, the overall difference between {\it sideR} and {\it sideL} $d_i$ values is considerably reduced when going from $\lambda =$ 5 to 10 \AA\, and for $\lambda= 20$ \AA\, the  $d_i$ distribution becomes almost uniform. This indicates that the characteristic correlation distances involved in the breathing mode are within the range of 5 to 20 \AA (see SI).
    
    \section{Conclusions}
    
    In the present work we have introduced a strategy based on the eigenvector centrality (EC) and mutual information metrics as a way of elucidating the allosteric pathways at an atomistic level and disentangle the local and non-local components of the characteristic distances that determine the allosteric mechanism. Furthermore, we have introduced a new perspective to measure centrality in terms of the environment relevance, allowing to interpret recent site directed mutagenesis experiments \citep{PNAS-Lisi-2017}.
    
    As opposed to other principal component analysis of widespread use in the literature, the EC scheme presented in this work provides a way to obtain the major collective correlation mode, independent from the magnitude of the fluctuations. As a consequence, this methodology constitutes a powerful strategy to quantify the relevance of each amino acid in the overall pathways of momentum transfer. In addition, the correlation measure is based on the generalized mutual information, which correctly captures the non-collinear correlation, overcoming the well known limitation of the Pearson correlation coefficients.
    
    We have used the IGPS protein as a test case to show that our approach successfully predicts the most important residues involved in the allosteric mechanism upon effector binding. The identified amino acids are localized around {\it sideR} of the {\it HisH}-{\it HisF} interface connecting the effector and the active sites. These residues belong to the same allosteric pathways detected in our previous community network analysis \cite{Rivalta2012} further corroborated by recent experimental evidences \cite{PNAS-Lisi-2017}. The outcome indicates how the comparative EC analysis here developed can predict allosteric pathways and estimate the role of long-range correlations in allostery through a robust and cost effective protocol.
    
    \section{acknowledgement}
    We acknowledge support by the Office of Basic Energy Sciences of the U.S. Department of Energy (DE-FG02-07ER15909) for supercomputer time from NERSC, XSEDE and the Yale University Faculty of Arts and Sciences High Performance Computing Center partially funded by the National Science Foundation grant CNS 08-21132. 
    
    \bibliography{allostery.bib}

\begin{thebibliography}{46}%
\makeatletter
\providecommand \@ifxundefined [1]{%
 \@ifx{#1\undefined}
}%
\providecommand \@ifnum [1]{%
 \ifnum #1\expandafter \@firstoftwo
 \else \expandafter \@secondoftwo
 \fi
}%
\providecommand \@ifx [1]{%
 \ifx #1\expandafter \@firstoftwo
 \else \expandafter \@secondoftwo
 \fi
}%
\providecommand \natexlab [1]{#1}%
\providecommand \enquote  [1]{``#1''}%
\providecommand \bibnamefont  [1]{#1}%
\providecommand \bibfnamefont [1]{#1}%
\providecommand \citenamefont [1]{#1}%
\providecommand \href@noop [0]{\@secondoftwo}%
\providecommand \href [0]{\begingroup \@sanitize@url \@href}%
\providecommand \@href[1]{\@@startlink{#1}\@@href}%
\providecommand \@@href[1]{\endgroup#1\@@endlink}%
\providecommand \@sanitize@url [0]{\catcode `\\12\catcode `\$12\catcode
  `\&12\catcode `\#12\catcode `\^12\catcode `\_12\catcode `\%12\relax}%
\providecommand \@@startlink[1]{}%
\providecommand \@@endlink[0]{}%
\providecommand \url  [0]{\begingroup\@sanitize@url \@url }%
\providecommand \@url [1]{\endgroup\@href {#1}{\urlprefix }}%
\providecommand \urlprefix  [0]{URL }%
\providecommand \Eprint [0]{\href }%
\providecommand \doibase [0]{http://dx.doi.org/}%
\providecommand \selectlanguage [0]{\@gobble}%
\providecommand \bibinfo  [0]{\@secondoftwo}%
\providecommand \bibfield  [0]{\@secondoftwo}%
\providecommand \translation [1]{[#1]}%
\providecommand \BibitemOpen [0]{}%
\providecommand \bibitemStop [0]{}%
\providecommand \bibitemNoStop [0]{.\EOS\space}%
\providecommand \EOS [0]{\spacefactor3000\relax}%
\providecommand \BibitemShut  [1]{\csname bibitem#1\endcsname}%
\let\auto@bib@innerbib\@empty
\bibitem [{\citenamefont {Csermely}\ \emph {et~al.}(2013)\citenamefont
  {Csermely}, \citenamefont {Korcsm{\'{a}}ros}, \citenamefont {Kiss},
  \citenamefont {London},\ and\ \citenamefont {Nussinov}}]{Csermely2013}%
  \BibitemOpen
  \bibfield  {author} {\bibinfo {author} {\bibfnamefont {P.}~\bibnamefont
  {Csermely}}, \bibinfo {author} {\bibfnamefont {T.}~\bibnamefont
  {Korcsm{\'{a}}ros}}, \bibinfo {author} {\bibfnamefont {H.~J.~M.}\
  \bibnamefont {Kiss}}, \bibinfo {author} {\bibfnamefont {G.}~\bibnamefont
  {London}}, \ and\ \bibinfo {author} {\bibfnamefont {R.}~\bibnamefont
  {Nussinov}},\ }\href {\doibase 10.1016/j.pharmthera.2013.01.016} {\bibfield
  {journal} {\bibinfo  {journal} {Pharmacology and Therapeutics}\ }\textbf
  {\bibinfo {volume} {138}},\ \bibinfo {pages} {333} (\bibinfo {year}
  {2013})},\ \Eprint {http://arxiv.org/abs/1210.0330} {arXiv:1210.0330}
  \BibitemShut {NoStop}%
\bibitem [{\citenamefont {Wagner}\ \emph {et~al.}(2016)\citenamefont {Wagner},
  \citenamefont {Lee}, \citenamefont {Durrant}, \citenamefont {Malmstrom},
  \citenamefont {Feher},\ and\ \citenamefont {Amaro}}]{Wagner2016}%
  \BibitemOpen
  \bibfield  {author} {\bibinfo {author} {\bibfnamefont {J.~R.}\ \bibnamefont
  {Wagner}}, \bibinfo {author} {\bibfnamefont {C.~T.}\ \bibnamefont {Lee}},
  \bibinfo {author} {\bibfnamefont {J.~D.}\ \bibnamefont {Durrant}}, \bibinfo
  {author} {\bibfnamefont {R.~D.}\ \bibnamefont {Malmstrom}}, \bibinfo {author}
  {\bibfnamefont {V.~A.}\ \bibnamefont {Feher}}, \ and\ \bibinfo {author}
  {\bibfnamefont {R.~E.}\ \bibnamefont {Amaro}},\ }\href {\doibase
  10.1021/acs.chemrev.5b00631} {\bibfield  {journal} {\bibinfo  {journal}
  {Chemical Reviews}\ }\textbf {\bibinfo {volume} {116}},\ \bibinfo {pages}
  {6370} (\bibinfo {year} {2016})}\BibitemShut {NoStop}%
\bibitem [{\citenamefont {Goodey}\ and\ \citenamefont
  {Benkovic}(2008)}]{Goodey2008}%
  \BibitemOpen
  \bibfield  {author} {\bibinfo {author} {\bibfnamefont {N.~M.}\ \bibnamefont
  {Goodey}}\ and\ \bibinfo {author} {\bibfnamefont {S.~J.}\ \bibnamefont
  {Benkovic}},\ }\href@noop {} {\bibfield  {journal} {\bibinfo  {journal} {Nat.
  Chem. Biol.}\ }\textbf {\bibinfo {volume} {4}},\ \bibinfo {pages} {478 }
  (\bibinfo {year} {2008})}\BibitemShut {NoStop}%
\bibitem [{\citenamefont {Reetz}\ \emph {et~al.}(2009)\citenamefont {Reetz},
  \citenamefont {Soni}, \citenamefont {Acevedo},\ and\ \citenamefont
  {Sanchis}}]{Reetz2009}%
  \BibitemOpen
  \bibfield  {author} {\bibinfo {author} {\bibfnamefont {M.~T.}\ \bibnamefont
  {Reetz}}, \bibinfo {author} {\bibfnamefont {P.}~\bibnamefont {Soni}},
  \bibinfo {author} {\bibfnamefont {J.~P.}\ \bibnamefont {Acevedo}}, \ and\
  \bibinfo {author} {\bibfnamefont {J.}~\bibnamefont {Sanchis}},\ }\href@noop
  {} {\bibfield  {journal} {\bibinfo  {journal} {Angew. Chem}\ }\textbf
  {\bibinfo {volume} {121}},\ \bibinfo {pages} {8418} (\bibinfo {year}
  {2009})}\BibitemShut {NoStop}%
\bibitem [{\citenamefont {Ozbil}\ \emph {et~al.}(2012)\citenamefont {Ozbil},
  \citenamefont {Barman}, \citenamefont {Bora},\ and\ \citenamefont
  {Prabhakar}}]{Ozbil2012}%
  \BibitemOpen
  \bibfield  {author} {\bibinfo {author} {\bibfnamefont {M.}~\bibnamefont
  {Ozbil}}, \bibinfo {author} {\bibfnamefont {A.}~\bibnamefont {Barman}},
  \bibinfo {author} {\bibfnamefont {R.~P.}\ \bibnamefont {Bora}}, \ and\
  \bibinfo {author} {\bibfnamefont {R.}~\bibnamefont {Prabhakar}},\ }\href
  {\doibase 10.1021/jz301597k} {\bibfield  {journal} {\bibinfo  {journal} {J.
  Phys. Chem. Lett.}\ }\textbf {\bibinfo {volume} {3}},\ \bibinfo {pages}
  {3460} (\bibinfo {year} {2012})}\BibitemShut {NoStop}%
\bibitem [{\citenamefont {Hawkins}\ and\ \citenamefont
  {McLeish}(2004)}]{Hawkins2004}%
  \BibitemOpen
  \bibfield  {author} {\bibinfo {author} {\bibfnamefont {R.~J.}\ \bibnamefont
  {Hawkins}}\ and\ \bibinfo {author} {\bibfnamefont {T.~C.~B.}\ \bibnamefont
  {McLeish}},\ }\href {\doibase 10.1103/PhysRevLett.93.098104} {\bibfield
  {journal} {\bibinfo  {journal} {Phys. Rev. Lett.}\ }\textbf {\bibinfo
  {volume} {93}},\ \bibinfo {pages} {098104} (\bibinfo {year}
  {2004})}\BibitemShut {NoStop}%
\bibitem [{\citenamefont {Ming}\ and\ \citenamefont {Wall}(2005)}]{Ming2005}%
  \BibitemOpen
  \bibfield  {author} {\bibinfo {author} {\bibfnamefont {D.}~\bibnamefont
  {Ming}}\ and\ \bibinfo {author} {\bibfnamefont {M.~E.}\ \bibnamefont
  {Wall}},\ }\href {\doibase 10.1103/PhysRevLett.95.198103} {\bibfield
  {journal} {\bibinfo  {journal} {Phys. Rev. Lett.}\ }\textbf {\bibinfo
  {volume} {95}},\ \bibinfo {pages} {198103} (\bibinfo {year}
  {2005})}\BibitemShut {NoStop}%
\bibitem [{\citenamefont {Palumbo}\ \emph {et~al.}(2006)\citenamefont
  {Palumbo}, \citenamefont {Farina}, \citenamefont {Colosimo}, \citenamefont
  {Tun},\ and\ \citenamefont {Dhar}}]{Palumbo2006}%
  \BibitemOpen
  \bibfield  {author} {\bibinfo {author} {\bibfnamefont {M.}~\bibnamefont
  {Palumbo}}, \bibinfo {author} {\bibfnamefont {L.}~\bibnamefont {Farina}},
  \bibinfo {author} {\bibfnamefont {A.}~\bibnamefont {Colosimo}}, \bibinfo
  {author} {\bibfnamefont {K.}~\bibnamefont {Tun}}, \ and\ \bibinfo {author}
  {\bibfnamefont {P.~K.}\ \bibnamefont {Dhar}},\ }\href@noop {} {\bibfield
  {journal} {\bibinfo  {journal} {Curr. Bioinf.}\ }\textbf {\bibinfo {volume}
  {1}},\ \bibinfo {pages} {219} (\bibinfo {year} {2006})}\BibitemShut {NoStop}%
\bibitem [{\citenamefont {Rivalta}\ \emph
  {et~al.}(2012{\natexlab{a}})\citenamefont {Rivalta}, \citenamefont {Sultan},
  \citenamefont {Lee}, \citenamefont {Manley}, \citenamefont {Loria},\ and\
  \citenamefont {Batista}}]{Rivalta2012}%
  \BibitemOpen
  \bibfield  {author} {\bibinfo {author} {\bibfnamefont {I.}~\bibnamefont
  {Rivalta}}, \bibinfo {author} {\bibfnamefont {M.~M.}\ \bibnamefont {Sultan}},
  \bibinfo {author} {\bibfnamefont {N.-S.}\ \bibnamefont {Lee}}, \bibinfo
  {author} {\bibfnamefont {G.~A.}\ \bibnamefont {Manley}}, \bibinfo {author}
  {\bibfnamefont {J.~P.}\ \bibnamefont {Loria}}, \ and\ \bibinfo {author}
  {\bibfnamefont {V.~S.}\ \bibnamefont {Batista}},\ }\href {\doibase
  10.1073/pnas.1120536109} {\bibfield  {journal} {\bibinfo  {journal} {Proc.
  Natl. Acad. Sci. USA}\ }\textbf {\bibinfo {volume} {109}},\ \bibinfo {pages}
  {E1428} (\bibinfo {year} {2012}{\natexlab{a}})},\ \Eprint
  {http://arxiv.org/abs/http://www.pnas.org/content/109/22/E1428.full.pdf+html}
  {http://www.pnas.org/content/109/22/E1428.full.pdf+html} \BibitemShut
  {NoStop}%
\bibitem [{\citenamefont {Vanwart}\ \emph {et~al.}(2012)\citenamefont
  {Vanwart}, \citenamefont {Eargle}, \citenamefont {Luthey-Schulten},\ and\
  \citenamefont {Amaro}}]{VanWart2012}%
  \BibitemOpen
  \bibfield  {author} {\bibinfo {author} {\bibfnamefont {A.~T.}\ \bibnamefont
  {Vanwart}}, \bibinfo {author} {\bibfnamefont {J.}~\bibnamefont {Eargle}},
  \bibinfo {author} {\bibfnamefont {Z.}~\bibnamefont {Luthey-Schulten}}, \ and\
  \bibinfo {author} {\bibfnamefont {R.~E.}\ \bibnamefont {Amaro}},\ }\href
  {\doibase 10.1021/ct300377a} {\bibfield  {journal} {\bibinfo  {journal}
  {Journal of Chemical Theory and Computation}\ }\textbf {\bibinfo {volume}
  {8}},\ \bibinfo {pages} {2949} (\bibinfo {year} {2012})},\ \Eprint
  {http://arxiv.org/abs/NIHMS150003} {arXiv:NIHMS150003} \BibitemShut {NoStop}%
\bibitem [{\citenamefont {Ribeiro}\ and\ \citenamefont
  {Ortiz}(2016)}]{Ribeiro2016}%
  \BibitemOpen
  \bibfield  {author} {\bibinfo {author} {\bibfnamefont {A.~A. S.~T.}\
  \bibnamefont {Ribeiro}}\ and\ \bibinfo {author} {\bibfnamefont
  {V.}~\bibnamefont {Ortiz}},\ }\href {\doibase 10.1021/acs.chemrev.5b00543}
  {\bibfield  {journal} {\bibinfo  {journal} {Chemical Reviews}\ }\textbf
  {\bibinfo {volume} {116}},\ \bibinfo {pages} {6488} (\bibinfo {year}
  {2016})}\BibitemShut {NoStop}%
\bibitem [{\citenamefont {Blacklock}\ and\ \citenamefont
  {Verkhivker}(2014)}]{Plos-Blacklock-2014}%
  \BibitemOpen
  \bibfield  {author} {\bibinfo {author} {\bibfnamefont {K.}~\bibnamefont
  {Blacklock}}\ and\ \bibinfo {author} {\bibfnamefont {G.~M.}\ \bibnamefont
  {Verkhivker}},\ }\href {https://doi.org/10.1371/journal.pcbi.1003679}
  {\bibfield  {journal} {\bibinfo  {journal} {PLOS Computational Biology}\
  }\textbf {\bibinfo {volume} {10}},\ \bibinfo {pages} {1} (\bibinfo {year}
  {2014})}\BibitemShut {NoStop}%
\bibitem [{\citenamefont {Sun}\ \emph {et~al.}(2014)\citenamefont {Sun},
  \citenamefont {Ågren},\ and\ \citenamefont {Tu}}]{JPCB-Yaoquan-2014}%
  \BibitemOpen
  \bibfield  {author} {\bibinfo {author} {\bibfnamefont {X.}~\bibnamefont
  {Sun}}, \bibinfo {author} {\bibfnamefont {H.}~\bibnamefont {Ågren}}, \ and\
  \bibinfo {author} {\bibfnamefont {Y.}~\bibnamefont {Tu}},\ }\href {\doibase
  10.1021/jp506579a} {\bibfield  {journal} {\bibinfo  {journal} {The Journal of
  Physical Chemistry B}\ }\textbf {\bibinfo {volume} {118}},\ \bibinfo {pages}
  {14737} (\bibinfo {year} {2014})},\ \bibinfo {note} {pMID: 25453446},\
  \Eprint {http://arxiv.org/abs/http://dx.doi.org/10.1021/jp506579a}
  {http://dx.doi.org/10.1021/jp506579a} \BibitemShut {NoStop}%
\bibitem [{\citenamefont {Zhu}\ \emph {et~al.}(2016)\citenamefont {Zhu},
  \citenamefont {Ma}, \citenamefont {Qi}, \citenamefont {Nussinov},\ and\
  \citenamefont {Zhang}}]{JPCB-Yuzhen-2016}%
  \BibitemOpen
  \bibfield  {author} {\bibinfo {author} {\bibfnamefont {Y.}~\bibnamefont
  {Zhu}}, \bibinfo {author} {\bibfnamefont {B.}~\bibnamefont {Ma}}, \bibinfo
  {author} {\bibfnamefont {R.}~\bibnamefont {Qi}}, \bibinfo {author}
  {\bibfnamefont {R.}~\bibnamefont {Nussinov}}, \ and\ \bibinfo {author}
  {\bibfnamefont {Q.}~\bibnamefont {Zhang}},\ }\href {\doibase
  10.1021/acs.jpcb.5b12299} {\bibfield  {journal} {\bibinfo  {journal} {The
  Journal of Physical Chemistry B}\ }\textbf {\bibinfo {volume} {120}},\
  \bibinfo {pages} {3551} (\bibinfo {year} {2016})},\ \bibinfo {note} {pMID:
  27007011},\ \Eprint
  {http://arxiv.org/abs/http://dx.doi.org/10.1021/acs.jpcb.5b12299}
  {http://dx.doi.org/10.1021/acs.jpcb.5b12299} \BibitemShut {NoStop}%
\bibitem [{\citenamefont {Appadurai}\ and\ \citenamefont
  {Senapati}(2016)}]{Biochem-sanjib-2016}%
  \BibitemOpen
  \bibfield  {author} {\bibinfo {author} {\bibfnamefont {R.}~\bibnamefont
  {Appadurai}}\ and\ \bibinfo {author} {\bibfnamefont {S.}~\bibnamefont
  {Senapati}},\ }\href {\doibase 10.1021/acs.biochem.5b00946} {\bibfield
  {journal} {\bibinfo  {journal} {Biochemistry}\ }\textbf {\bibinfo {volume}
  {55}},\ \bibinfo {pages} {1529} (\bibinfo {year} {2016})},\ \bibinfo {note}
  {pMID: 26892689},\ \Eprint
  {http://arxiv.org/abs/http://dx.doi.org/10.1021/acs.biochem.5b00946}
  {http://dx.doi.org/10.1021/acs.biochem.5b00946} \BibitemShut {NoStop}%
\bibitem [{\citenamefont {Xu}\ \emph {et~al.}(2015)\citenamefont {Xu},
  \citenamefont {Ye}, \citenamefont {Jiang}, \citenamefont {Yang},
  \citenamefont {Zhang}, \citenamefont {Feng}, \citenamefont {Luo},\ and\
  \citenamefont {Chen}}]{JPCB-Lishi-2015}%
  \BibitemOpen
  \bibfield  {author} {\bibinfo {author} {\bibfnamefont {L.}~\bibnamefont
  {Xu}}, \bibinfo {author} {\bibfnamefont {W.}~\bibnamefont {Ye}}, \bibinfo
  {author} {\bibfnamefont {C.}~\bibnamefont {Jiang}}, \bibinfo {author}
  {\bibfnamefont {J.}~\bibnamefont {Yang}}, \bibinfo {author} {\bibfnamefont
  {J.}~\bibnamefont {Zhang}}, \bibinfo {author} {\bibfnamefont
  {Y.}~\bibnamefont {Feng}}, \bibinfo {author} {\bibfnamefont {R.}~\bibnamefont
  {Luo}}, \ and\ \bibinfo {author} {\bibfnamefont {H.-F.}\ \bibnamefont
  {Chen}},\ }\href {\doibase 10.1021/jp510940w} {\bibfield  {journal} {\bibinfo
   {journal} {The Journal of Physical Chemistry B}\ }\textbf {\bibinfo {volume}
  {119}},\ \bibinfo {pages} {2844} (\bibinfo {year} {2015})},\ \bibinfo {note}
  {pMID: 25633018},\ \Eprint
  {http://arxiv.org/abs/http://dx.doi.org/10.1021/jp510940w}
  {http://dx.doi.org/10.1021/jp510940w} \BibitemShut {NoStop}%
\bibitem [{\citenamefont {VanWart}\ \emph {et~al.}(2012)\citenamefont
  {VanWart}, \citenamefont {Eargle}, \citenamefont {Luthey-Schulten},\ and\
  \citenamefont {Amaro}}]{JCTC-VanWart-2012}%
  \BibitemOpen
  \bibfield  {author} {\bibinfo {author} {\bibfnamefont {A.~T.}\ \bibnamefont
  {VanWart}}, \bibinfo {author} {\bibfnamefont {J.}~\bibnamefont {Eargle}},
  \bibinfo {author} {\bibfnamefont {Z.}~\bibnamefont {Luthey-Schulten}}, \ and\
  \bibinfo {author} {\bibfnamefont {R.~E.}\ \bibnamefont {Amaro}},\ }\href
  {\doibase 10.1021/ct300377a} {\bibfield  {journal} {\bibinfo  {journal}
  {Journal of Chemical Theory and Computation}\ }\textbf {\bibinfo {volume}
  {8}},\ \bibinfo {pages} {2949} (\bibinfo {year} {2012})},\ \bibinfo {note}
  {pMID: 23139645},\ \Eprint
  {http://arxiv.org/abs/http://dx.doi.org/10.1021/ct300377a}
  {http://dx.doi.org/10.1021/ct300377a} \BibitemShut {NoStop}%
\bibitem [{\citenamefont {Palermo}\ \emph {et~al.}(0)\citenamefont {Palermo},
  \citenamefont {Ricci}, \citenamefont {Fernando}, \citenamefont {Basak},
  \citenamefont {Jinek}, \citenamefont {Rivalta}, \citenamefont {Batista},\
  and\ \citenamefont {McCammon}}]{JACS-palermo-2017}%
  \BibitemOpen
  \bibfield  {author} {\bibinfo {author} {\bibfnamefont {G.}~\bibnamefont
  {Palermo}}, \bibinfo {author} {\bibfnamefont {C.~G.}\ \bibnamefont {Ricci}},
  \bibinfo {author} {\bibfnamefont {A.}~\bibnamefont {Fernando}}, \bibinfo
  {author} {\bibfnamefont {R.}~\bibnamefont {Basak}}, \bibinfo {author}
  {\bibfnamefont {M.}~\bibnamefont {Jinek}}, \bibinfo {author} {\bibfnamefont
  {I.}~\bibnamefont {Rivalta}}, \bibinfo {author} {\bibfnamefont {V.~S.}\
  \bibnamefont {Batista}}, \ and\ \bibinfo {author} {\bibfnamefont {J.~A.}\
  \bibnamefont {McCammon}},\ }\href {\doibase 10.1021/jacs.7b05313} {\bibfield
  {journal} {\bibinfo  {journal} {Journal of the American Chemical Society}\
  }\textbf {\bibinfo {volume} {0}},\ \bibinfo {pages} {null} (\bibinfo {year}
  {0})},\ \bibinfo {note} {pMID: 28764328},\ \Eprint
  {http://arxiv.org/abs/http://dx.doi.org/10.1021/jacs.7b05313}
  {http://dx.doi.org/10.1021/jacs.7b05313} \BibitemShut {NoStop}%
\bibitem [{\citenamefont {Guo}\ and\ \citenamefont
  {Zhou}(2016)}]{Chemrev-Guo-2016}%
  \BibitemOpen
  \bibfield  {author} {\bibinfo {author} {\bibfnamefont {J.}~\bibnamefont
  {Guo}}\ and\ \bibinfo {author} {\bibfnamefont {H.-X.}\ \bibnamefont {Zhou}},\
  }\href {\doibase 10.1021/acs.chemrev.5b00590} {\bibfield  {journal} {\bibinfo
   {journal} {Chemical Reviews}\ }\textbf {\bibinfo {volume} {116}},\ \bibinfo
  {pages} {6503} (\bibinfo {year} {2016})},\ \bibinfo {note} {pMID: 26876046},\
  \Eprint {http://arxiv.org/abs/http://dx.doi.org/10.1021/acs.chemrev.5b00590}
  {http://dx.doi.org/10.1021/acs.chemrev.5b00590} \BibitemShut {NoStop}%
\bibitem [{\citenamefont {Li}\ \emph {et~al.}(2014)\citenamefont {Li},
  \citenamefont {Zhang}, \citenamefont {Lu}, \citenamefont {Huang},
  \citenamefont {Geng}, \citenamefont {Shen},\ and\ \citenamefont
  {Zhang}}]{PONE-Li-2014}%
  \BibitemOpen
  \bibfield  {author} {\bibinfo {author} {\bibfnamefont {S.}~\bibnamefont
  {Li}}, \bibinfo {author} {\bibfnamefont {J.}~\bibnamefont {Zhang}}, \bibinfo
  {author} {\bibfnamefont {S.}~\bibnamefont {Lu}}, \bibinfo {author}
  {\bibfnamefont {W.}~\bibnamefont {Huang}}, \bibinfo {author} {\bibfnamefont
  {L.}~\bibnamefont {Geng}}, \bibinfo {author} {\bibfnamefont {Q.}~\bibnamefont
  {Shen}}, \ and\ \bibinfo {author} {\bibfnamefont {J.}~\bibnamefont {Zhang}},\
  }\href {https://doi.org/10.1371/journal.pone.0097668} {\bibfield  {journal}
  {\bibinfo  {journal} {PLOS ONE}\ }\textbf {\bibinfo {volume} {9}},\ \bibinfo
  {pages} {1} (\bibinfo {year} {2014})}\BibitemShut {NoStop}%
\bibitem [{\citenamefont {Sethi}\ \emph {et~al.}(2009)\citenamefont {Sethi},
  \citenamefont {Eargle}, \citenamefont {Black},\ and\ \citenamefont
  {Luthey-Schulten}}]{PNAS-Sethi-2009}%
  \BibitemOpen
  \bibfield  {author} {\bibinfo {author} {\bibfnamefont {A.}~\bibnamefont
  {Sethi}}, \bibinfo {author} {\bibfnamefont {J.}~\bibnamefont {Eargle}},
  \bibinfo {author} {\bibfnamefont {A.~A.}\ \bibnamefont {Black}}, \ and\
  \bibinfo {author} {\bibfnamefont {Z.}~\bibnamefont {Luthey-Schulten}},\
  }\href {\doibase 10.1073/pnas.0810961106} {\bibfield  {journal} {\bibinfo
  {journal} {Proceedings of the National Academy of Sciences}\ }\textbf
  {\bibinfo {volume} {106}},\ \bibinfo {pages} {6620} (\bibinfo {year}
  {2009})}\BibitemShut {NoStop}%
\bibitem [{\citenamefont {Ricci}\ \emph {et~al.}(2016)\citenamefont {Ricci},
  \citenamefont {Silveira}, \citenamefont {Rivalta}, \citenamefont {Batista},\
  and\ \citenamefont {Skaf}}]{SCIREP-Ricci-2016}%
  \BibitemOpen
  \bibfield  {author} {\bibinfo {author} {\bibfnamefont {C.~G.}\ \bibnamefont
  {Ricci}}, \bibinfo {author} {\bibfnamefont {R.~L.}\ \bibnamefont {Silveira}},
  \bibinfo {author} {\bibfnamefont {I.}~\bibnamefont {Rivalta}}, \bibinfo
  {author} {\bibfnamefont {V.~S.}\ \bibnamefont {Batista}}, \ and\ \bibinfo
  {author} {\bibfnamefont {M.~S.}\ \bibnamefont {Skaf}},\ }\href@noop {}
  {\bibfield  {journal} {\bibinfo  {journal} {Scientific Reports}\ }\textbf
  {\bibinfo {volume} {6}},\ \bibinfo {pages} {19940} (\bibinfo {year}
  {2016})}\BibitemShut {NoStop}%
\bibitem [{\citenamefont {Papaleo}\ \emph {et~al.}(2012)\citenamefont
  {Papaleo}, \citenamefont {Lindorff-Larsen},\ and\ \citenamefont
  {De~Gioia}}]{Papaleo2012}%
  \BibitemOpen
  \bibfield  {author} {\bibinfo {author} {\bibfnamefont {E.}~\bibnamefont
  {Papaleo}}, \bibinfo {author} {\bibfnamefont {K.}~\bibnamefont
  {Lindorff-Larsen}}, \ and\ \bibinfo {author} {\bibfnamefont {L.}~\bibnamefont
  {De~Gioia}},\ }\href@noop {} {\bibfield  {journal} {\bibinfo  {journal}
  {Phys. Chem. Chem. Phys.}\ }\textbf {\bibinfo {volume} {14}},\ \bibinfo
  {pages} {12515} (\bibinfo {year} {2012})}\BibitemShut {NoStop}%
\bibitem [{\citenamefont {David-Eden}\ and\ \citenamefont
  {Mandel-Gufreund}(2008)}]{David-Eden2008}%
  \BibitemOpen
  \bibfield  {author} {\bibinfo {author} {\bibfnamefont {H.}~\bibnamefont
  {David-Eden}}\ and\ \bibinfo {author} {\bibfnamefont {Y.}~\bibnamefont
  {Mandel-Gufreund}},\ }\href@noop {} {\bibfield  {journal} {\bibinfo
  {journal} {Nucleic Acid Res.}\ }\textbf {\bibinfo {volume} {36}},\ \bibinfo
  {pages} {4641} (\bibinfo {year} {2008})}\BibitemShut {NoStop}%
\bibitem [{\citenamefont {Jiang}\ \emph {et~al.}(2010)\citenamefont {Jiang},
  \citenamefont {Chen},\ and\ \citenamefont {Xiao}}]{Jiang2010}%
  \BibitemOpen
  \bibfield  {author} {\bibinfo {author} {\bibfnamefont {X.}~\bibnamefont
  {Jiang}}, \bibinfo {author} {\bibfnamefont {C.}~\bibnamefont {Chen}}, \ and\
  \bibinfo {author} {\bibfnamefont {Y.}~\bibnamefont {Xiao}},\ }\href@noop {}
  {\bibfield  {journal} {\bibinfo  {journal} {J. Comput. Chem.}\ }\textbf
  {\bibinfo {volume} {31}},\ \bibinfo {pages} {2502} (\bibinfo {year}
  {2010})}\BibitemShut {NoStop}%
\bibitem [{\citenamefont {Szilagyi}\ \emph {et~al.}(2013)\citenamefont
  {Szilagyi}, \citenamefont {Nussinov},\ and\ \citenamefont
  {Csermely}}]{Szilagyi2013}%
  \BibitemOpen
  \bibfield  {author} {\bibinfo {author} {\bibfnamefont {A.}~\bibnamefont
  {Szilagyi}}, \bibinfo {author} {\bibfnamefont {R.}~\bibnamefont {Nussinov}},
  \ and\ \bibinfo {author} {\bibfnamefont {P.}~\bibnamefont {Csermely}},\
  }\href@noop {} {\bibfield  {journal} {\bibinfo  {journal} {Curr. Topics in
  Med. Chem.}\ }\textbf {\bibinfo {volume} {13}},\ \bibinfo {pages} {64}
  (\bibinfo {year} {2013})}\BibitemShut {NoStop}%
\bibitem [{\citenamefont {Lange}\ and\ \citenamefont
  {Grubm\"{u}lller}(2006)}]{Lange2006}%
  \BibitemOpen
  \bibfield  {author} {\bibinfo {author} {\bibfnamefont {O.}~\bibnamefont
  {Lange}}\ and\ \bibinfo {author} {\bibfnamefont {H.}~\bibnamefont
  {Grubm\"{u}lller}},\ }\href@noop {} {\bibfield  {journal} {\bibinfo
  {journal} {Proteins: Structure, Function, and Bioinformatics}\ }\textbf
  {\bibinfo {volume} {62}},\ \bibinfo {pages} {1053} (\bibinfo {year}
  {2006})}\BibitemShut {NoStop}%
\bibitem [{\citenamefont {Lange}\ and\ \citenamefont
  {Grubmüller}(2008)}]{Lange-2008}%
  \BibitemOpen
  \bibfield  {author} {\bibinfo {author} {\bibfnamefont {O.~F.}\ \bibnamefont
  {Lange}}\ and\ \bibinfo {author} {\bibfnamefont {H.}~\bibnamefont
  {Grubmüller}},\ }\href {\doibase 10.1002/prot.21618} {\bibfield  {journal}
  {\bibinfo  {journal} {Proteins: Structure, Function, and Bioinformatics}\
  }\textbf {\bibinfo {volume} {70}},\ \bibinfo {pages} {1294} (\bibinfo {year}
  {2008})}\BibitemShut {NoStop}%
\bibitem [{\citenamefont {Savoie}\ \emph {et~al.}(2014)\citenamefont {Savoie},
  \citenamefont {Kohlstedt}, \citenamefont {Jackson}, \citenamefont {Chen},
  \citenamefont {Olvera de~la Cruz}, \citenamefont {Schatz}, \citenamefont
  {Marks},\ and\ \citenamefont {Ratner}}]{Savoie-2014}%
  \BibitemOpen
  \bibfield  {author} {\bibinfo {author} {\bibfnamefont {B.~M.}\ \bibnamefont
  {Savoie}}, \bibinfo {author} {\bibfnamefont {K.~L.}\ \bibnamefont
  {Kohlstedt}}, \bibinfo {author} {\bibfnamefont {N.~E.}\ \bibnamefont
  {Jackson}}, \bibinfo {author} {\bibfnamefont {L.~X.}\ \bibnamefont {Chen}},
  \bibinfo {author} {\bibfnamefont {M.}~\bibnamefont {Olvera de~la Cruz}},
  \bibinfo {author} {\bibfnamefont {G.~C.}\ \bibnamefont {Schatz}}, \bibinfo
  {author} {\bibfnamefont {T.~J.}\ \bibnamefont {Marks}}, \ and\ \bibinfo
  {author} {\bibfnamefont {M.~A.}\ \bibnamefont {Ratner}},\ }\href {\doibase
  10.1073/pnas.1409514111} {\bibfield  {journal} {\bibinfo  {journal}
  {Proceedings of the National Academy of Sciences}\ }\textbf {\bibinfo
  {volume} {111}},\ \bibinfo {pages} {10055} (\bibinfo {year} {2014})},\
  \Eprint
  {http://arxiv.org/abs/http://www.pnas.org/content/111/28/10055.full.pdf}
  {http://www.pnas.org/content/111/28/10055.full.pdf} \BibitemShut {NoStop}%
\bibitem [{\citenamefont {Doshi}\ \emph {et~al.}(2016)\citenamefont {Doshi},
  \citenamefont {Holliday}, \citenamefont {Eisenmesser},\ and\ \citenamefont
  {Hamelberg}}]{Doshi-2016}%
  \BibitemOpen
  \bibfield  {author} {\bibinfo {author} {\bibfnamefont {U.}~\bibnamefont
  {Doshi}}, \bibinfo {author} {\bibfnamefont {M.~J.}\ \bibnamefont {Holliday}},
  \bibinfo {author} {\bibfnamefont {E.~Z.}\ \bibnamefont {Eisenmesser}}, \ and\
  \bibinfo {author} {\bibfnamefont {D.}~\bibnamefont {Hamelberg}},\ }\href
  {\doibase 10.1073/pnas.1523573113} {\bibfield  {journal} {\bibinfo  {journal}
  {Proceedings of the National Academy of Sciences}\ }\textbf {\bibinfo
  {volume} {113}},\ \bibinfo {pages} {4735} (\bibinfo {year} {2016})},\ \Eprint
  {http://arxiv.org/abs/http://www.pnas.org/content/113/17/4735.full.pdf}
  {http://www.pnas.org/content/113/17/4735.full.pdf} \BibitemShut {NoStop}%
\bibitem [{\citenamefont {Ruhnau}(2000)}]{Ruhnau2000}%
  \BibitemOpen
  \bibfield  {author} {\bibinfo {author} {\bibfnamefont {B.}~\bibnamefont
  {Ruhnau}},\ }\href {\doibase 10.1016/S0378-8733(00)00031-9} {\bibfield
  {journal} {\bibinfo  {journal} {Social Networks}\ }\textbf {\bibinfo {volume}
  {22}},\ \bibinfo {pages} {357} (\bibinfo {year} {2000})}\BibitemShut
  {NoStop}%
\bibitem [{\citenamefont {Chaudhuri}\ \emph {et~al.}(2001)\citenamefont
  {Chaudhuri}, \citenamefont {Lange}, \citenamefont {Myers}, \citenamefont
  {Chittur}, \citenamefont {Davisson},\ and\ \citenamefont
  {Smith}}]{Chaudhuri-2016}%
  \BibitemOpen
  \bibfield  {author} {\bibinfo {author} {\bibfnamefont {B.~N.}\ \bibnamefont
  {Chaudhuri}}, \bibinfo {author} {\bibfnamefont {S.~C.}\ \bibnamefont
  {Lange}}, \bibinfo {author} {\bibfnamefont {R.~S.}\ \bibnamefont {Myers}},
  \bibinfo {author} {\bibfnamefont {S.~V.}\ \bibnamefont {Chittur}}, \bibinfo
  {author} {\bibfnamefont {V.}~\bibnamefont {Davisson}}, \ and\ \bibinfo
  {author} {\bibfnamefont {J.~L.}\ \bibnamefont {Smith}},\ }\href@noop {}
  {\bibfield  {journal} {\bibinfo  {journal} {Structure}\ }\textbf {\bibinfo
  {volume} {9}},\ \bibinfo {pages} {987} (\bibinfo {year} {2001})}\BibitemShut
  {NoStop}%
\bibitem [{\citenamefont {Myers}\ \emph {et~al.}(2003)\citenamefont {Myers},
  \citenamefont {Jensen}, \citenamefont {Deras}, \citenamefont {Smith},\ and\
  \citenamefont {Davisson}}]{Myers2003}%
  \BibitemOpen
  \bibfield  {author} {\bibinfo {author} {\bibfnamefont {R.~S.}\ \bibnamefont
  {Myers}}, \bibinfo {author} {\bibfnamefont {J.~R.}\ \bibnamefont {Jensen}},
  \bibinfo {author} {\bibfnamefont {I.~L.}\ \bibnamefont {Deras}}, \bibinfo
  {author} {\bibfnamefont {J.~L.}\ \bibnamefont {Smith}}, \ and\ \bibinfo
  {author} {\bibfnamefont {V.~J.}\ \bibnamefont {Davisson}},\ }\href {\doibase
  10.1021/bi034314l} {\bibfield  {journal} {\bibinfo  {journal} {Biochem.}\
  }\textbf {\bibinfo {volume} {42}},\ \bibinfo {pages} {7013} (\bibinfo {year}
  {2003})}\BibitemShut {NoStop}%
\bibitem [{\citenamefont {Rivalta}\ \emph
  {et~al.}(2012{\natexlab{b}})\citenamefont {Rivalta}, \citenamefont {Sultan},
  \citenamefont {Lee}, \citenamefont {Manley}, \citenamefont {Loria},\ and\
  \citenamefont {Batista}}]{Rivalta2012SI}%
  \BibitemOpen
  \bibfield  {author} {\bibinfo {author} {\bibfnamefont {I.}~\bibnamefont
  {Rivalta}}, \bibinfo {author} {\bibfnamefont {M.~M.}\ \bibnamefont {Sultan}},
  \bibinfo {author} {\bibfnamefont {N.-S.}\ \bibnamefont {Lee}}, \bibinfo
  {author} {\bibfnamefont {G.~A.}\ \bibnamefont {Manley}}, \bibinfo {author}
  {\bibfnamefont {J.~P.}\ \bibnamefont {Loria}}, \ and\ \bibinfo {author}
  {\bibfnamefont {V.~S.}\ \bibnamefont {Batista}},\ }\href {\doibase
  10.1073/pnas.1120536109} {\bibfield  {journal} {\bibinfo  {journal} {Proc.
  Natl. Acad. Sci. USA}\ }\textbf {\bibinfo {volume} {109}},\ \bibinfo {pages}
  {E1428} (\bibinfo {year} {2012}{\natexlab{b}})},\ \bibinfo {note} {supporting
  information doccument},\ \Eprint
  {http://arxiv.org/abs/http://www.pnas.org/content/109/22/E1428.full.pdf+html}
  {http://www.pnas.org/content/109/22/E1428.full.pdf+html} \BibitemShut
  {NoStop}%
\bibitem [{\citenamefont {Floyd}(1962)}]{Floyd1962}%
  \BibitemOpen
  \bibfield  {author} {\bibinfo {author} {\bibfnamefont {R.~W.}\ \bibnamefont
  {Floyd}},\ }\href@noop {} {\bibfield  {journal} {\bibinfo  {journal} {Commun.
  Acm.}\ }\textbf {\bibinfo {volume} {5}},\ \bibinfo {pages} {345} (\bibinfo
  {year} {1962})}\BibitemShut {NoStop}%
\bibitem [{\citenamefont {Girvan}\ and\ \citenamefont
  {Newman}(2002)}]{Girvan2002}%
  \BibitemOpen
  \bibfield  {author} {\bibinfo {author} {\bibfnamefont {M.}~\bibnamefont
  {Girvan}}\ and\ \bibinfo {author} {\bibfnamefont {M.~E.~J.}\ \bibnamefont
  {Newman}},\ }\href {\doibase 10.1073/pnas.122653799} {\bibfield  {journal}
  {\bibinfo  {journal} {Proc. Natl. Acad. Sci. USA}\ }\textbf {\bibinfo
  {volume} {99}},\ \bibinfo {pages} {7821} (\bibinfo {year} {2002})},\ \Eprint
  {http://arxiv.org/abs/http://www.pnas.org/content/99/12/7821.full.pdf+html}
  {http://www.pnas.org/content/99/12/7821.full.pdf+html} \BibitemShut {NoStop}%
\bibitem [{\citenamefont {Newman}(2006)}]{Newman2006}%
  \BibitemOpen
  \bibfield  {author} {\bibinfo {author} {\bibfnamefont {M.~E.~J.}\
  \bibnamefont {Newman}},\ }\href {\doibase 10.1073/pnas.0601602103} {\bibfield
   {journal} {\bibinfo  {journal} {Proceedings of the National Academy of
  Sciences of the United States of America}\ }\textbf {\bibinfo {volume}
  {103}},\ \bibinfo {pages} {8577} (\bibinfo {year} {2006})},\ \Eprint
  {http://arxiv.org/abs/0602124} {arXiv:0602124 [physics]} \BibitemShut
  {NoStop}%
\bibitem [{\citenamefont {Rivalta}\ \emph {et~al.}(2016)\citenamefont
  {Rivalta}, \citenamefont {Lisi}, \citenamefont {Snoeberger}, \citenamefont
  {Manley}, \citenamefont {Loria},\ and\ \citenamefont
  {Batista}}]{Rivalta2016}%
  \BibitemOpen
  \bibfield  {author} {\bibinfo {author} {\bibfnamefont {I.}~\bibnamefont
  {Rivalta}}, \bibinfo {author} {\bibfnamefont {G.~P.}\ \bibnamefont {Lisi}},
  \bibinfo {author} {\bibfnamefont {N.-S.}\ \bibnamefont {Snoeberger}},
  \bibinfo {author} {\bibfnamefont {G.}~\bibnamefont {Manley}}, \bibinfo
  {author} {\bibfnamefont {J.~P.}\ \bibnamefont {Loria}}, \ and\ \bibinfo
  {author} {\bibfnamefont {V.~S.}\ \bibnamefont {Batista}},\ }\href {\doibase
  10.1021/acs.biochem.6b00859} {\bibfield  {journal} {\bibinfo  {journal}
  {Biochem.}\ }\textbf {\bibinfo {volume} {55}},\ \bibinfo {pages} {6484}
  (\bibinfo {year} {2016})},\ \bibinfo {note} {pMID: 27797506},\ \Eprint
  {http://arxiv.org/abs/http://dx.doi.org/10.1021/acs.biochem.6b00859}
  {http://dx.doi.org/10.1021/acs.biochem.6b00859} \BibitemShut {NoStop}%
\bibitem [{\citenamefont {Lisi}\ \emph {et~al.}(2017)\citenamefont {Lisi},
  \citenamefont {East}, \citenamefont {Batista},\ and\ \citenamefont
  {Loria}}]{PNAS-Lisi-2017}%
  \BibitemOpen
  \bibfield  {author} {\bibinfo {author} {\bibfnamefont {G.~P.}\ \bibnamefont
  {Lisi}}, \bibinfo {author} {\bibfnamefont {K.~W.}\ \bibnamefont {East}},
  \bibinfo {author} {\bibfnamefont {V.~S.}\ \bibnamefont {Batista}}, \ and\
  \bibinfo {author} {\bibfnamefont {J.~P.}\ \bibnamefont {Loria}},\ }\href
  {\doibase 10.1073/pnas.1700448114} {\bibfield  {journal} {\bibinfo  {journal}
  {Proceedings of the National Academy of Sciences}\ }\textbf {\bibinfo
  {volume} {114}},\ \bibinfo {pages} {E3414} (\bibinfo {year} {2017})},\
  \Eprint
  {http://arxiv.org/abs/http://www.pnas.org/content/114/17/E3414.full.pdf}
  {http://www.pnas.org/content/114/17/E3414.full.pdf} \BibitemShut {NoStop}%
\bibitem [{\citenamefont {Watkins}(2010)}]{watkins2010}%
  \BibitemOpen
  \bibfield  {author} {\bibinfo {author} {\bibfnamefont {D.~S.}\ \bibnamefont
  {Watkins}},\ }\href@noop {} {\emph {\bibinfo {title} {Fundamentals of matrix
  computations, third edition}}}\ (\bibinfo  {publisher} {John Wiley \& Sons},\
  \bibinfo {year} {2010})\BibitemShut {NoStop}%
\bibitem [{\citenamefont {Jimenez-Martinez}\ and\ \citenamefont
  {Negre}(2017)}]{Jimenez2017}%
  \BibitemOpen
  \bibfield  {author} {\bibinfo {author} {\bibfnamefont {J.}~\bibnamefont
  {Jimenez-Martinez}}\ and\ \bibinfo {author} {\bibfnamefont {C.~F.~A.}\
  \bibnamefont {Negre}},\ }\href {\doibase 10.1103/PhysRevE.96.013310}
  {\bibfield  {journal} {\bibinfo  {journal} {Phys. Rev. E}\ }\textbf {\bibinfo
  {volume} {96}},\ \bibinfo {pages} {013310} (\bibinfo {year}
  {2017})}\BibitemShut {NoStop}%
\bibitem [{\citenamefont {Lisi}\ \emph {et~al.}(2016)\citenamefont {Lisi},
  \citenamefont {Manley}, \citenamefont {Hendrickson}, \citenamefont {Rivalta},
  \citenamefont {Batista},\ and\ \citenamefont {Loria}}]{Lisi-struct-2016}%
  \BibitemOpen
  \bibfield  {author} {\bibinfo {author} {\bibfnamefont {G.}~\bibnamefont
  {Lisi}}, \bibinfo {author} {\bibfnamefont {G.}~\bibnamefont {Manley}},
  \bibinfo {author} {\bibfnamefont {H.}~\bibnamefont {Hendrickson}}, \bibinfo
  {author} {\bibfnamefont {I.}~\bibnamefont {Rivalta}}, \bibinfo {author}
  {\bibfnamefont {V.~S.}\ \bibnamefont {Batista}}, \ and\ \bibinfo {author}
  {\bibfnamefont {J.}~\bibnamefont {Loria}},\ }\href {\doibase
  https://doi.org/10.1016/j.str.2016.04.010} {\bibfield  {journal} {\bibinfo
  {journal} {Structure}\ }\textbf {\bibinfo {volume} {24}},\ \bibinfo {pages}
  {1155 } (\bibinfo {year} {2016})}\BibitemShut {NoStop}%
\bibitem [{\citenamefont {Amadei}\ \emph {et~al.}(1993)\citenamefont {Amadei},
  \citenamefont {Linssen},\ and\ \citenamefont {Berendsen}}]{ED-Amadei1993}%
  \BibitemOpen
  \bibfield  {author} {\bibinfo {author} {\bibfnamefont {A.}~\bibnamefont
  {Amadei}}, \bibinfo {author} {\bibfnamefont {A.~B.~M.}\ \bibnamefont
  {Linssen}}, \ and\ \bibinfo {author} {\bibfnamefont {H.~J.~C.}\ \bibnamefont
  {Berendsen}},\ }\href {\doibase 10.1002/prot.340170408} {\bibfield  {journal}
  {\bibinfo  {journal} {Proteins: Structure, Function, and Bioinformatics}\
  }\textbf {\bibinfo {volume} {17}},\ \bibinfo {pages} {412} (\bibinfo {year}
  {1993})}\BibitemShut {NoStop}%
\bibitem [{\citenamefont {Hayward}\ and\ \citenamefont
  {de~Groot}(2008)}]{ED-Hayward2008}%
  \BibitemOpen
  \bibfield  {author} {\bibinfo {author} {\bibfnamefont {S.}~\bibnamefont
  {Hayward}}\ and\ \bibinfo {author} {\bibfnamefont {B.~L.}\ \bibnamefont
  {de~Groot}},\ }\enquote {\bibinfo {title} {Normal modes and essential
  dynamics},}\ in\ \href {\doibase 10.1007/978-1-59745-177-2_5} {\emph
  {\bibinfo {booktitle} {Molecular Modeling of Proteins}}},\ \bibinfo {editor}
  {edited by\ \bibinfo {editor} {\bibfnamefont {A.}~\bibnamefont {Kukol}}}\
  (\bibinfo  {publisher} {Humana Press},\ \bibinfo {address} {Totowa, NJ},\
  \bibinfo {year} {2008})\ pp.\ \bibinfo {pages} {89--106}\BibitemShut
  {NoStop}%
\bibitem [{\citenamefont {Meyer}\ \emph {et~al.}(2006)\citenamefont {Meyer},
  \citenamefont {Ferrer-Costa}, \citenamefont {Pérez}, \citenamefont {Rueda},
  \citenamefont {Bidon-Chanal}, \citenamefont {Luque}, \citenamefont
  {Laughton},\ and\ \citenamefont {Orozco}}]{ED-meyer2006}%
  \BibitemOpen
  \bibfield  {author} {\bibinfo {author} {\bibfnamefont {T.}~\bibnamefont
  {Meyer}}, \bibinfo {author} {\bibfnamefont {C.}~\bibnamefont {Ferrer-Costa}},
  \bibinfo {author} {\bibfnamefont {A.}~\bibnamefont {Pérez}}, \bibinfo
  {author} {\bibfnamefont {M.}~\bibnamefont {Rueda}}, \bibinfo {author}
  {\bibfnamefont {A.}~\bibnamefont {Bidon-Chanal}}, \bibinfo {author}
  {\bibfnamefont {F.~J.}\ \bibnamefont {Luque}}, \bibinfo {author}
  {\bibfnamefont {C.~A.}\ \bibnamefont {Laughton}}, \ and\ \bibinfo {author}
  {\bibfnamefont {M.}~\bibnamefont {Orozco}},\ }\href {\doibase
  10.1021/ct050285b} {\bibfield  {journal} {\bibinfo  {journal} {Journal of
  Chemical Theory and Computation}\ }\textbf {\bibinfo {volume} {2}},\ \bibinfo
  {pages} {251} (\bibinfo {year} {2006})}\BibitemShut {NoStop}%
\bibitem [{\citenamefont {Morzan}\ \emph {et~al.}(2013)\citenamefont {Morzan},
  \citenamefont {Capece}, \citenamefont {Marti},\ and\ \citenamefont
  {Estrin}}]{ED-prot2013}%
  \BibitemOpen
  \bibfield  {author} {\bibinfo {author} {\bibfnamefont {U.~N.}\ \bibnamefont
  {Morzan}}, \bibinfo {author} {\bibfnamefont {L.}~\bibnamefont {Capece}},
  \bibinfo {author} {\bibfnamefont {M.~A.}\ \bibnamefont {Marti}}, \ and\
  \bibinfo {author} {\bibfnamefont {D.~A.}\ \bibnamefont {Estrin}},\ }\href
  {\doibase 10.1002/prot.24245} {\bibfield  {journal} {\bibinfo  {journal}
  {Proteins: Structure, Function, and Bioinformatics}\ }\textbf {\bibinfo
  {volume} {81}},\ \bibinfo {pages} {863} (\bibinfo {year} {2013})}\BibitemShut
  {NoStop}%
\end{thebibliography}%
    
\end{document}